\documentclass[a4paper,fleqn,usenatbib]{mnras}

\usepackage{txfonts}

\usepackage[T1]{fontenc}
\usepackage{ae,aecompl}

\usepackage{psfig}   
\usepackage{graphicx}
\usepackage{amssymb}
\usepackage{subfigure}

\title[Spatial distribution of massive stars]{No preferential spatial distribution for massive stars expected from their formation}

\author[R. J. Parker \& J. E. Dale]{Richard  J. Parker$^{1}$\thanks{E-mail: R.Parker@sheffield.ac.uk}\thanks{Royal Society Dorothy Hodgkin Fellow} and James E. Dale$^{2}$ \vspace*{0.1cm}\\
$^{1}$Department of Physics and Astronomy, The University of Sheffield, Hicks Building, Hounsfield Road, Sheffield, S3 7RH, UK\\
$^{2}$Centre for Astrophysics Research, Science and Technology Research Institute, University of Hertfordshire, Hatfield, AL10 9AB, UK}

\begin{document}

                             
\pagerange{\pageref{firstpage}--\pageref{lastpage}} \pubyear{2016}

\maketitle

\label{firstpage}

\begin{abstract}
We analyse $N$-body and Smoothed Particle Hydrodynamic (SPH) simulations of young star-forming regions to search for differences in the spatial distributions of massive stars compared to lower-mass stars. The competitive accretion theory of massive star formation posits that the most massive stars should sit in deeper potential wells than lower-mass stars. This may be observable in the relative surface density or spatial concentration of the most massive stars compared to other, lower-mass stars. Massive stars in cool--collapse $N$-body models do end up in significantly deeper potentials, and are mass segregated. However, in models of warm (expanding) star-forming regions, whilst the massive stars do come to be in deeper potentials than average stars, they are not mass segregated. In the purely hydrodynamical SPH simulations, the massive stars do come to reside in deeper potentials, which is due to their runaway growth. However, when photoionisation and stellar winds are implemented in the simulations, these feedback mechanisms regulate the mass of the stars and disrupt the inflow of gas into the clouds' potential wells. This generally makes the potential wells shallower than in the control runs, and prevents the massive stars from occupying deeper potentials. This in turn results in the most massive stars having a very similar spatial concentration and surface density distribution to lower-mass stars. Whilst massive stars do form via competitive accretion in our simulations, this rarely translates to a different spatial distribution and so any lack of primordial mass segregation in an observed star-forming region does not preclude competitive accretion as a viable formation mechanism for massive stars.   
\end{abstract}


\section{Introduction}

Massive stars ($>$20\,M$_\odot$) play a crucial role in astrophysics. They are the dominant component of the luminosity of young star-forming regions and star clusters, and their high core temperatures, powerful winds and inability to support themselves through electron degeneracy pressure leads to the chemical enrichment of the Universe through their collective stellar evolution and explosions as supernovae.

Despite their importance, theories that explain the formation of massive stars in star-forming environments have not yet converged on a common formation channel \citep{Tan14}, with two very different formation mechanisms proposed in the literature  \citep[see the review by][]{Zinnecker07}. Competitive accretion \citep{Zinnecker82,Bonnell97,Bonnell98b,Bonnell01,Bonnell04} posits that all stars are initially Jeans-mass objects, and those that have access to larger gas reservoirs then grow more massive. On the other hand, monolithic collapse predicts a more top-down formation scenario \citep{McKee03,Krumholz05} and suggests that the mass of the prestellar core is the determining factor in how massive a star will eventually become.


One possible way to determine which (if either) of the monolithic collapse and competitive accretion scenarios is the dominant channel of massive star formation is to look for differences in the spatial distributions of the massive stars with respect to lower-mass stars. In particular, competitive accretion posits that the most massive stars grow from low-mass seeds due to their preferential position in the potential and so they can accrete more gas \citep{Bonnell97}. A prediction of the competitive accretion theory is therefore that the most massive stars should sit deeper in the potential well of their host protocluster or star--forming region. Interestingly, for the monlithic collapse scenario to be efficient, the massive prestellar core must be centrally located within the gas clump \citep{McKee03}, which would also predict some degree of primordial mass segregation \citep[see the discussion in section 8 of][]{Moeckel09a}.

If massive stars do sit in a deeper potential, then we might expect that once formed they will be more spatially concentrated than the average--mass star. However, no studies have shown whether there is a direct correlation between potential and spatial distribution.

Furthermore, observations are inconclusive as to whether massive stars do form with a different spatial distribution due to difficulties in observing massive star (and concurrent low-mass star) formation, although this is changing rapidly due the advent of observational facilities such as ALMA \citep[e.g.][]{Henshaw17}. 

Mass segregation -- where the most massive stars are more spatially concentrated than average stars -- has been observed in some star-forming regions and clusters with ages 1 -- 10\,Myr \citep{Allison09a,Pang13} but not in others \citep{Wright14,Gennaro17,Parker17a}. Furthermore, several authors have shown that dynamical mass segregation occurs rapidly \citep{McMillan07,Allison09b,Moeckel09b,Allison10,Olczak11,Parker14b,Parker16c}, which could account for the observed clusters that do show mass segregation. 

The lack of mass segregation in some star-forming regions and clusters, and the ease with which a region can dynamically segregate, suggests that the early stages of massive star formation does not lead to, nor require the massive stars being in a different spatial configuration to the lower-mass stars. However, it is unclear whether this in turn rules out competitive accretion as being the dominant channel for high mass star formation.

Simulations can provide information on the earliest stages of star formation, and in this paper we examine five sets of SPH simulations from \citet{Dale14} to determine whether the most massive stars that form through competitive accretion (a) sit in a deeper potential, (b) are more spatially concentrated and (c) are in regions of higher surface densities, than the average stars. 

The paper is organised as follows. In Section~2 we describe the metrics we use to quantify the spatial distribution of massive stars in our simulations. In Section 3 we first describe the evolution of the spatial distribution of massive stars for pure $N$-body simulations of star-forming regions with different initial conditions (collapsing versus expanding). We then quantify the potential, spatial concentration and relative density of the most massive stars in SPH simulations immediately after star formation and then at various stages in their later dynamical evolution. We provide a discussion in Section~4 and we conclude in Section~5.

\section{Method}

In this Section we first describe the three metrics used to quantify the relative spatial distribution of the most massive stars in a star-forming region compared to lower-mass stars, before describing the $N$-body and SPH simulations used to follow the dynamical evolution of our simulated star-forming regions.

\subsection{Quantifying the spatial distributions of stars}

\subsubsection{The potential difference ratio, $\Phi_{\rm PDR}$}

To search for differences between the gravitational potential of massive stars and lower-mass stars in our simulations we determine the gravitational potential, $\Phi_j$, for each star in the simulation:
\begin{equation}
\Phi_j = -\sum{\frac{m_i}{r_{ij}}},
\end{equation}
where $m_i$ is the $i^{\rm th}$ star in the summation, and $r_{ij}$ is the distance to the $i^{\rm th}$ star. In an analagous method to the surface density -- mass distribution method \citep[][see below]{Maschberger11}, we plot $\Phi_j$ against $m_j$ for each star. We then compare the median potential of different subsets of stars using the potential difference ratio.

The potential difference ratio, PDR, is defined as:
\begin{equation}
\Phi_{\rm PDR} = \frac{\tilde{\Phi}_{10}}{\tilde{\Phi}_{\rm all}},
\end{equation}
where $\tilde{\Phi}_{10}$ is the median potential of the 10 most massive stars, and $\tilde{\Phi}_{\rm all}$ is the median potential of the entire cluster. If $\Phi_{\rm PDR} > 1$ then the most massive stars sit deeper in the potential than the average star and we quantify the significance of this by means of a Kolmogorov-Smirnov (KS) test on the cumulative distribution of the stars, ranked by their potentials, where we reject the hypothesis that the two subsets are drawn from the same underlying distribution if the KS p-value is less than 0.1. This methodology is analogous to that used to define the local surface density ratio, $\Sigma_{\rm LDR}$ (see below), but as we shall see in the following sections, the two metrics do not measure the same thing.

\subsubsection{The mass segregation ratio, $\Lambda_{\rm MSR}$}

A common method used to quantify the spatial distribution of massive stars in star-forming regions is to use the Minimum Spanning Tree (MST) approach \citep{Allison09a,Parker15b}. We define the `mass  segregation ratio' 
($\Lambda_{\rm MSR}$) as the ratio between the average MST pathlength of 10 randomly chosen stars in a star-forming region and 
and that of the 10 most massive stars:
\begin{equation}
\Lambda_{\rm MSR} = {\frac{\langle l_{\rm average} \rangle}{l_{\rm 10}}} ^{+ {\sigma_{\rm 5/6}}/{l_{\rm 10}}}_{- {\sigma_{\rm 1/6}}/{l_{\rm 10}}}.
\end{equation}
There is a dispersion  associated
with the average length of random MSTs, which is roughly Gaussian and
can be quantified as the standard deviation  of the lengths  $\langle
l_{\rm average} \rangle \pm \sigma_{\rm average}$. However, we
conservatively estimate the lower (upper) uncertainty  as the MST
length which lies 1/6 (5/6) of the way through an ordered list of all
the random lengths (corresponding to a 66 per cent deviation from  the
median value, $\langle l_{\rm average} \rangle$). This determination
prevents a single outlying object from heavily influencing the
uncertainty.

If $\Lambda_{\rm MSR} > 1$, then the most massive stars are more spacially concentrated than the average stars, and we designate this as significant if the lower error bar also exceeds unity. Note that $\Lambda_{\rm MSR}$ can be too sensitive in that it sometimes finds that random fluctuations in low-number distributions lead to  mass segregation according to our definition \citep{Parker15b}. \citet{Parker15b} suggest that any values of $\Lambda_{\rm MSR}$ less than 2 should be treated with caution, even if the error bars suggest they significantly deviate from unity. 

\subsubsection{The local surface density ratio, $\Sigma_{\rm LDR}$} 

In addition to quantifying whether the most massive stars are mass segregated, we also determine the relative local surface density of the most massive stars compared to lower-mass stars \citep{Maschberger11,Kupper11}. We define the local surface density around each star, $\Sigma$ as
\begin{equation}
\Sigma = \frac{N - 1}{\pi r_N^2},
\end{equation}
where $r_N$ is the distance to the $N^{\rm th}$ nearest neighbour, $N$.  We adopt $N = 10$ throughout this work. 

We divide the median $\Sigma$ for the ten most massive stars, $\tilde{\Sigma}_\mathrm{10}$ by the median value for all the stars $\tilde{\Sigma}_\mathrm{all}$ to define a `local density ratio', $\Sigma_{\rm LDR}$ \citep{Parker14b}:
\begin{equation}
\Sigma_{\rm LDR} = \frac{\tilde{\Sigma}_\mathrm{10}}{\tilde{\Sigma}_\mathrm{all}}.
\end{equation} 
If $\Sigma_{\rm LDR} > 1$ then the most massive stars are in areas of higher local surface density than the average star and we quantify the significance of this by means of a KS test on the cumulative distribution of the stars, ranked by their local surface densities, where we reject the hypothesis that the two subsets are drawn from the same underlying distribution if the KS p-value is less than 0.1.\\

Throughout the analysis in this paper, we have chosen to compare the measured diagnostics for the ten most massive stars to those measured for the entire star-forming region. First, we note that the total number of stars varies between simulations. Usually we have several hundred stars, but several of our simulations have fewer than 100 stars. \citet{Parker15b} show that robust quantification of mass segregation can be made by comparing the 10 most massive stars to the full distribution in star-forming regions with several hundred stars.

Each of our statistical measures of the significance of any preferential spatial distribution for the most massive stars are asking the data whether there is a clear departure from the distribution of stars for the entire mass range. Therefore, such departures are harder to detect in low-$N$ systems ($<$100 stars).  However, we choose not to vary the number of stars in the most massive subset as a function of the total number of stars in a region because dynamical mass segregation (when it occurs) ususally operates on more than the ten most massive stars \citep{Parker14b,Parker16c}, even in regions of 200 stars or fewer, so any mass segregation (primordial or dynamical) should be detectable in at least this size of subset.

\subsection{Simulations of star-forming regions}

\subsubsection{Pure $N$-body simulations}

We first examine the evolution of pure $N$-body simulations of star-forming regions to compare the evolution of the three metrics used to quantify the spatial distributions of massive stars compared to low-mass stars. These $N$-body simulations are taken from \citet{Parker14b} and we refer the interested reader to that paper for full details of the simulation set-up. We briefly summarise the simulations below.

We follow the evolution of two types of simulation, each containing $N = 1500$ stars in a substructured (fractal) distribution set up according to the prescription in \citet{Goodwin04a}. The fractal dimension is $D = 1.6$, corresponding to a high level of substructure, and the velocities are correlated on local scales \citep[see e.g.][]{Parker14b,Parker16b}. We draw masses from the formulation of the IMF presented in \citet{Maschberger13}, which has the form 
\begin{equation}
p(m) \propto \left(\frac{m}{\mu}\right)^{-\alpha}\left(1 + \left(\frac{m}{\mu}\right)^{1 - \alpha}\right)^{-\beta}
\label{imf}.
\end{equation}
Here, $\mu = 0.2$\,M$_\odot$ is the average stellar mass, $\alpha = 2.3$ is the \citet{Salpeter55} power-law exponent for higher mass stars, and $\beta = 1.4$ is used to describe the slope of the IMF for low-mass objects \citep*[which also deviates from the log-normal form;][]{Bastian10}. Finally, we sample from this IMF within the mass range $m_{\rm low} = 0.01$\,M$_\odot$ to $m_{\rm up} = 50$\,M$_\odot$.

The velocities of stars are then scaled to the desired virial ratio, $\alpha_{\rm vir} = T/|\Omega|$, where $T$ and $|\Omega|$ are the total kinetic energy and total potential energy of the stars, respectively. A star-forming region is in virial equilibrium if $\alpha_{\rm vir} = 0.5$. One set of simulations are subvirial ($\alpha_{\rm vir} = 0.3$), i.e.\,\,collapsing, and the other set are supervirial  ($\alpha_{\rm vir} = 1.5$), i.e.\,\,expanding.

The simulations are evolved for 10\,Myr using the  $4^{\rm th}$ order Hermite-scheme integrator \texttt{kira}  within the Starlab environment \citep[e.g.][]{Zwart99,Zwart01}. Stellar evolution is also implemented using the \texttt{SeBa} stellar evolution package in Starlab \citep{Zwart96,Zwart12}, which provides look-up tables for the evolution of stars according to the time dependent mass-radius relations  in \citet*{Eggleton89} and \citet{Tout96}. A summary of the simulations is given in Table~\ref{nbody_cluster_props}.


\subsubsection{SPH simulations with subsequent $N$-body evolution}

The SPH simulations are taken from \citet{Dale14} and the detailed set-up is described in those papers. Briefly, \citet{Dale14} evolve five SPH simulations of Giant Molecular Clouds with different mass, radius and initial virial ratio. Each simulation is forked, when a few massive stars have formed,  into a control run and a feedback run, in which the effects of photoionisation and stellar winds are included. The simulations are evolved for as close as numerically feasible to 3\,Myr after the formation of the first few massive stars. After this time, the first supernova explosions are expected, which are not yet modelled, so the SPH models terminate at this point.

After the SPH simulations are stopped, we remove any remaining gas and evolve the sink particles as a pure $N$-body calculation for a further 10\,Myr using the \texttt{kira} integrator \citep{Zwart99,Zwart01}, including the effects  of stellar evolution from look-up tables in the  \texttt{SeBa} package \citep{Zwart96,Zwart12}. We do not alter the velocities of the stars to compensate for the sudden removal of the gas potential. However, we note that preliminary tests in our earlier work \citep[][their section 4.1]{Parker13a} found minimal differences between simulations in which the velocities of the sink particles were rescaled after gas removal to be in global virial equilibrium before the $N$-body evolution commenced. 

Following the transition from SPH to $N$-body, we do not see the formation of close binaries, or the rapid ejection of stars. This is in part be due to the fact that the sink particles reside in stellar-dominated potentials long before the removal of the gas (i.e. the end of the SPH simulations) and therefore removing the gas somewhat artificially does not drastically influence the behaviour of the sink particles.

Furthermore, \citet{Parker13a} found that spatial structure of the sink particles is preserved during the transition from SPH to $N$-body, meaning that any primordial mass segregation is not erased simply due to removing the gas potential -- rather if it is erased it will be due to a combination of later dynamical and/or stellar evolution \citep[see also][]{Moeckel09a}. In a future paper we will investigate in detail the effects of different scalings for the sink particle velocities (e.g. cold, subvirial, virialised) on the long-term survival of the clusters. 

The $N$-body evolution of the sink particles for the five pairs of simulations is described in detail in \citet{Parker13a} and \citet{Parker15a} and we refer the interested reader to those papers. The simulation id, number of sink particles, total mass in sink particles and median sink particle mass are given in Table~\ref{sph_cluster_props}.

\section{Results}

In this section we first describe the evolution of the potential difference ratio ($\Phi_{\rm PDR}$), mass segregation ($\Lambda_{\rm MSR}$) and relative surface densities ($\Sigma_{\rm LDR}$) of massive stars in our pure $N$-body simulations, before discussing the evolution of the these metrics in our SPH simulations.

\subsection{Cool-collapse $N$-body simulation}

\begin{table*}
\caption[bf]{Summary of the results from the $N$-body simulations. The columns display the type of simulation, corresponding initial virial ratio, and the number of particles, $N$, then we show at 0, 2, and 10\,Myr whether the potential difference ratio $\Phi_{\rm PDR} >> 1$, the mass segregation ratio $\Lambda_{\rm MSR} >> 1$ and the local surface density ratio  $\Sigma_{\rm LDR} >> 1$. If these ratios do not significantly exceed unity, the entry is left blank.}
\begin{center}
\begin{tabular}{|c|c|c|c|c|c|c|}
\hline 
Type & $\alpha_{\rm vir}$ &  $N$ & Time (Myr) & $\Phi_{\rm PDR} >> 1$? & $\Lambda_{\rm MSR} >> 1$?  & $\Sigma_{\rm LDR} >> 1$? \\
\hline
 &  &  & 0 &  & &  \\
 Cool collapse & 0.3 & 1500 & 2 & + & + & + \\
 &  &  & 10 & + & +  & +\\
\hline
\hline
 &  &  & 0 &  & &  \\
 Warm expansion & 1.5 & 1500 & 2 & + & & + \\
 &  &  & 10 & + & & + \\
\hline
\end{tabular}
\end{center}
\label{nbody_cluster_props}
\end{table*}

\begin{figure*}
  \begin{center}
\setlength{\subfigcapskip}{10pt}

\hspace*{-1.5cm}\subfigure[0\,Myr]{\label{subvirial-a}\rotatebox{270}{\includegraphics[scale=0.28]{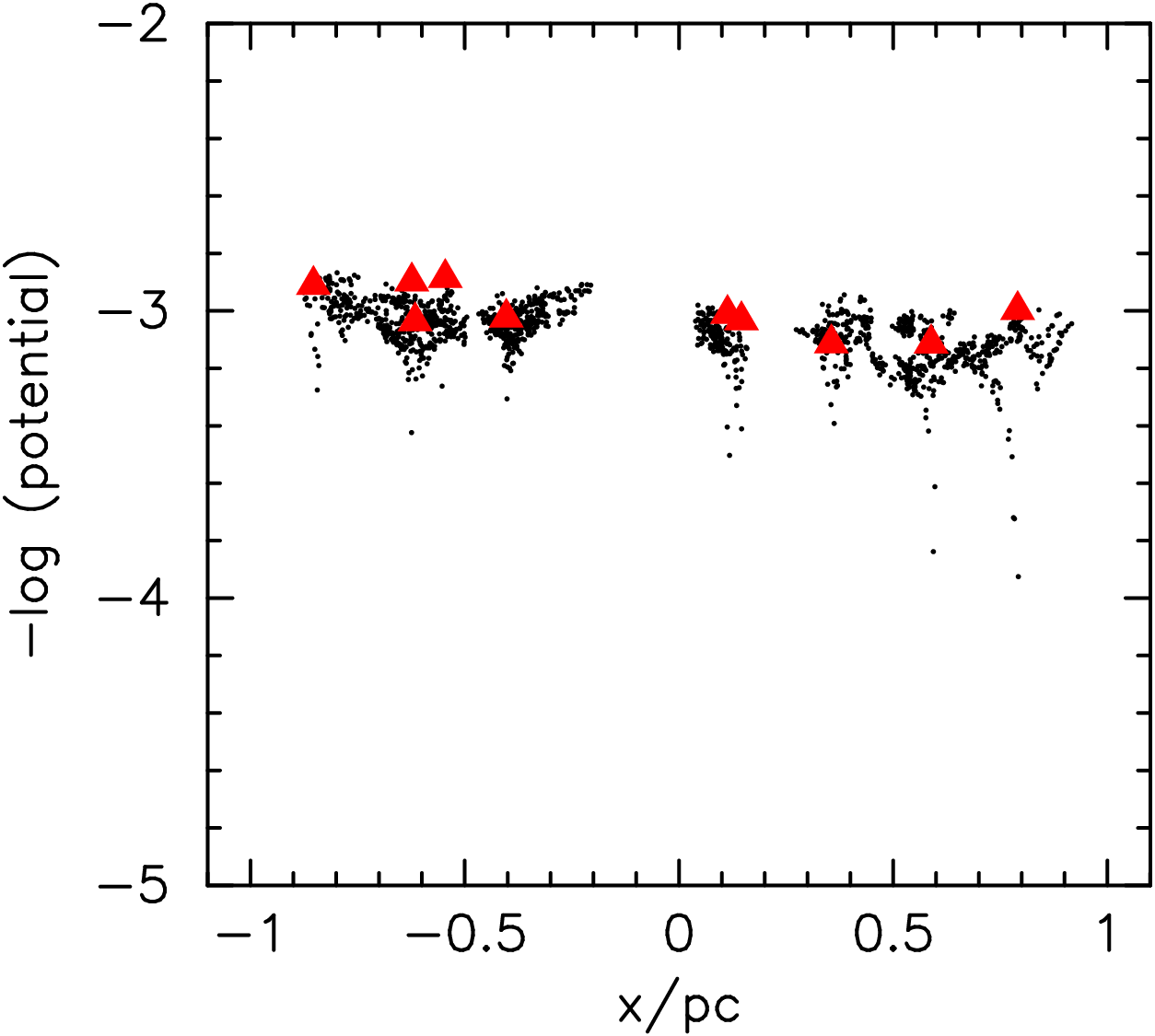}}}
\hspace*{0.3cm} 
\subfigure[4\,Myr]{\label{subvirial-b}\rotatebox{270}{\includegraphics[scale=0.28]{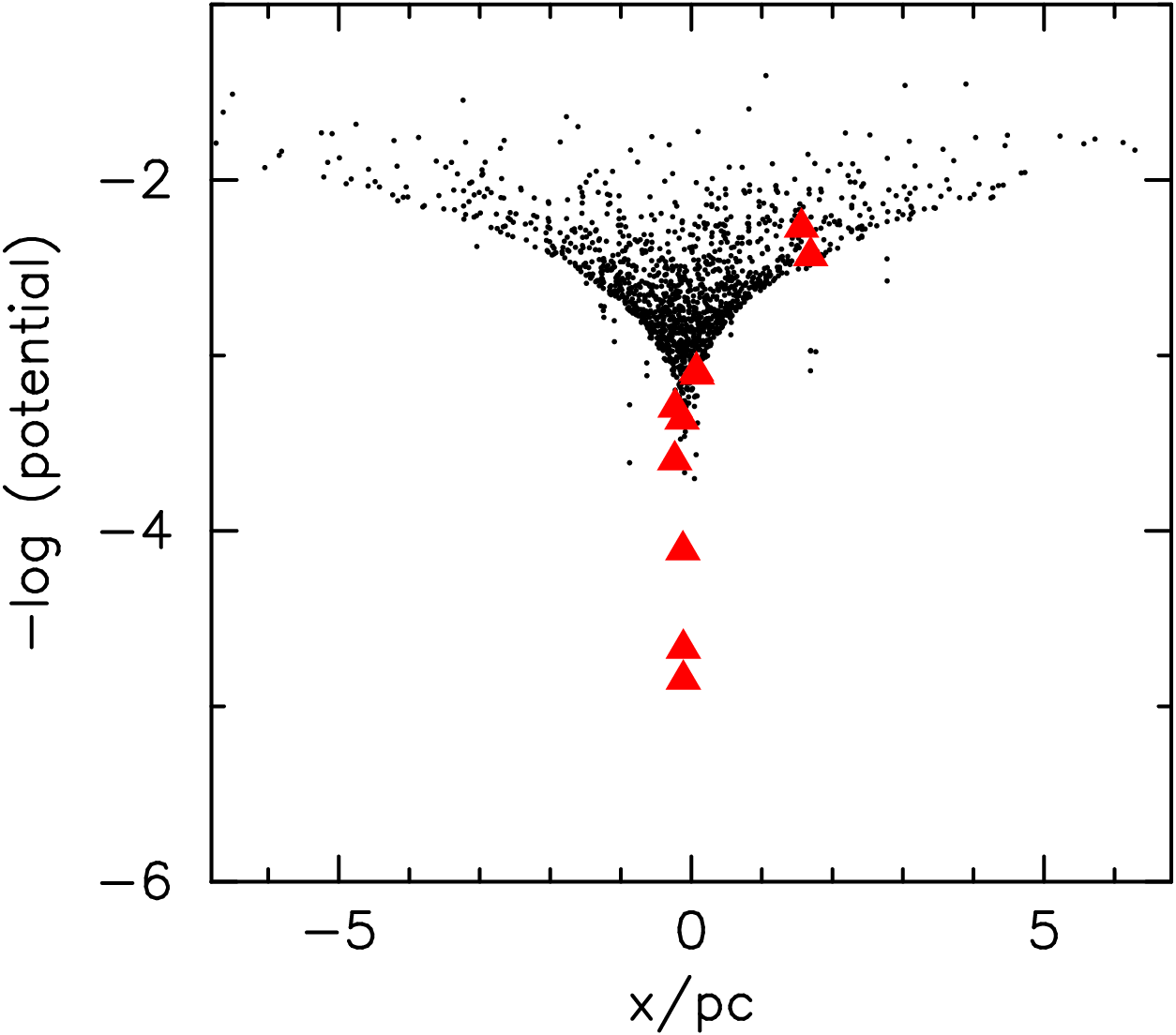}}} 
\hspace*{0.3cm}\subfigure[10\,Myr]{\label{subvirial-c}\rotatebox{270}{\includegraphics[scale=0.28]{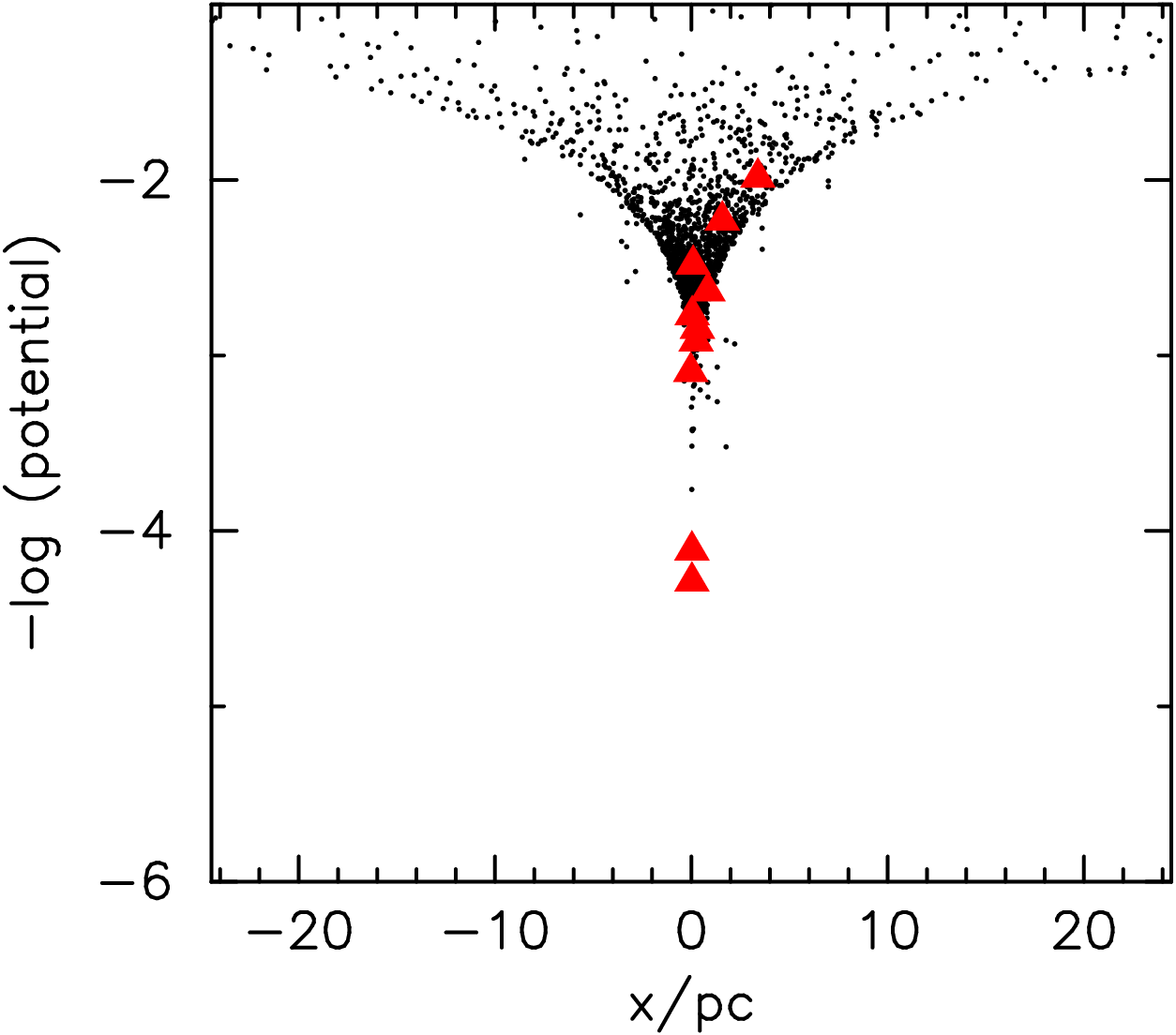}}}
\hspace*{-1.5cm}\subfigure[$\Phi - m$, 0\,Myr]{\label{subvirial-d}\rotatebox{270}{\includegraphics[scale=0.28]{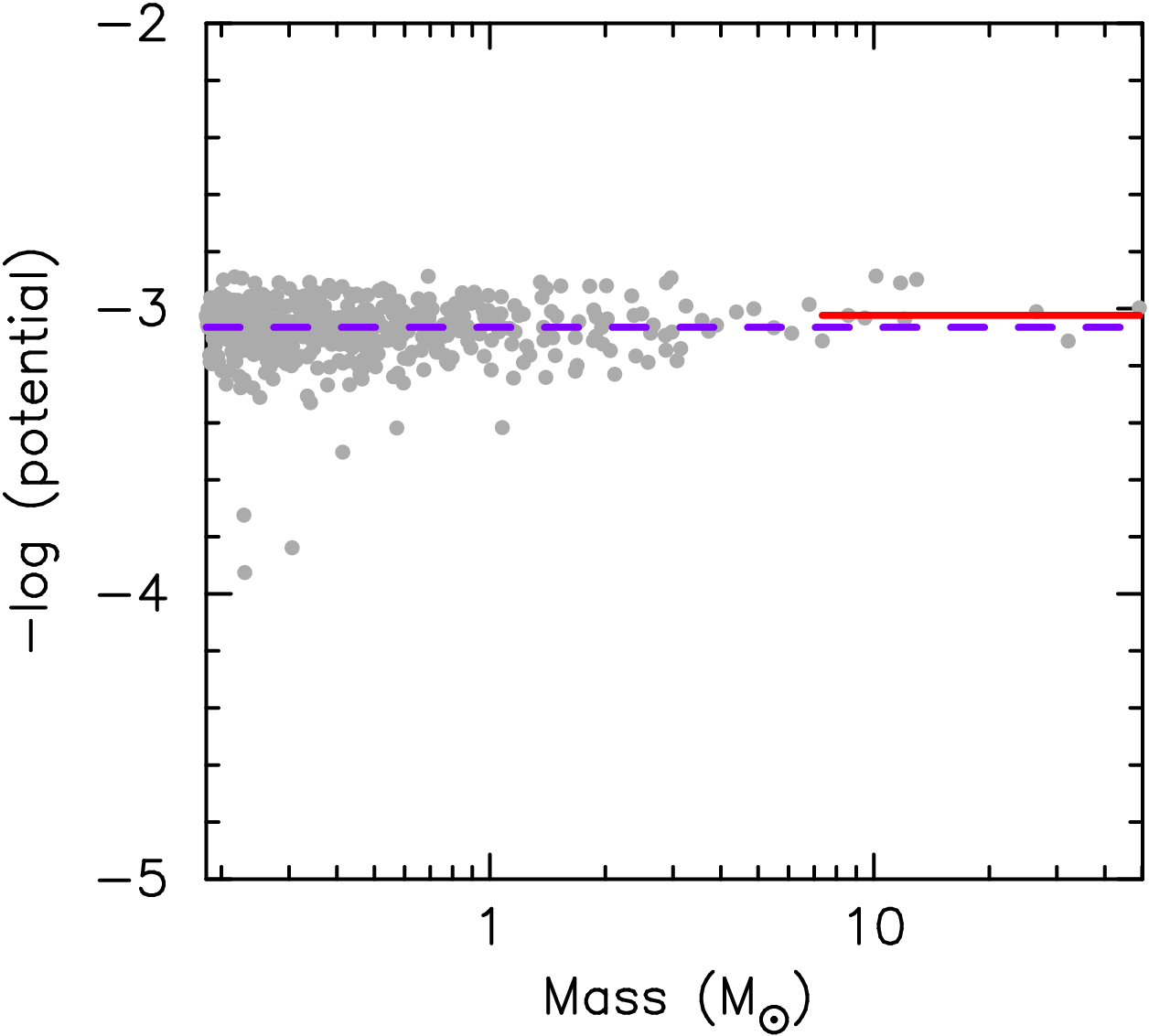}}}
\hspace*{0.3cm} 
\subfigure[$\Phi - m$, 4\,Myr]{\label{subvirial-e}\rotatebox{270}{\includegraphics[scale=0.28]{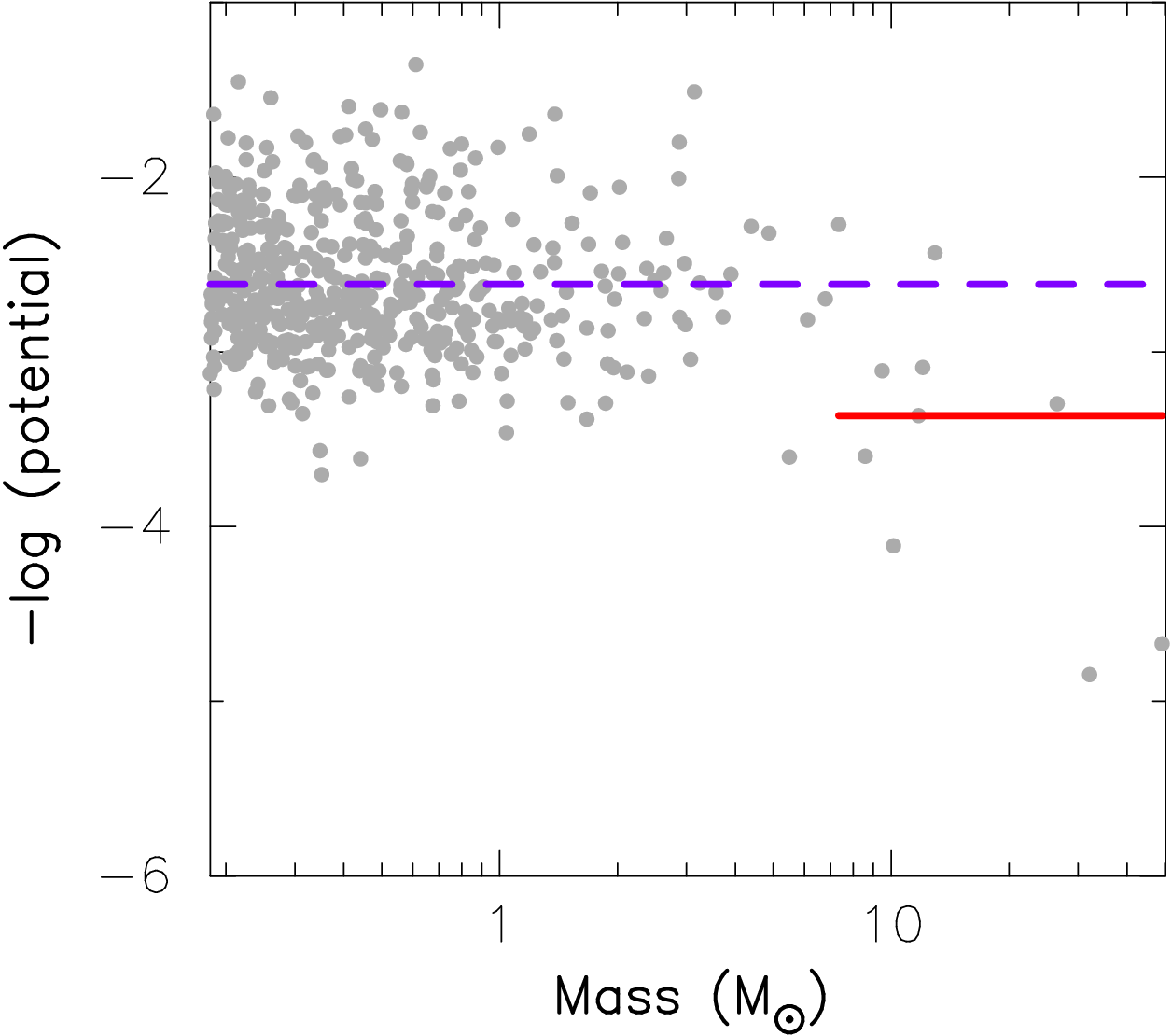}}} 
\hspace*{0.3cm}\subfigure[$\Phi - m$, 10\,Myr]{\label{subvirial-f}\rotatebox{270}{\includegraphics[scale=0.28]{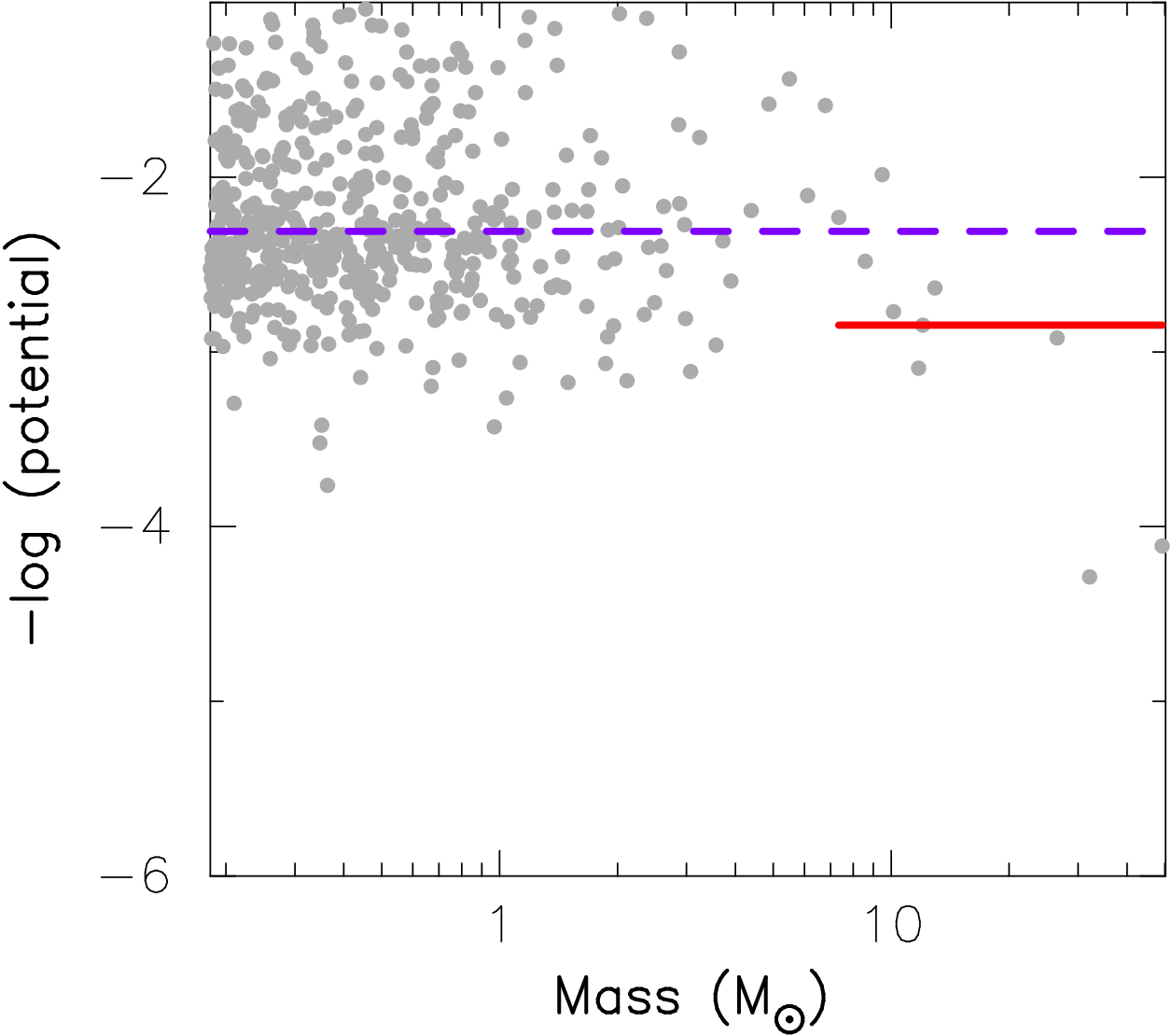}}}
\hspace*{-1.5cm}\subfigure[Evolution of the potential]{\label{subvirial-g}\rotatebox{270}{\includegraphics[scale=0.28]{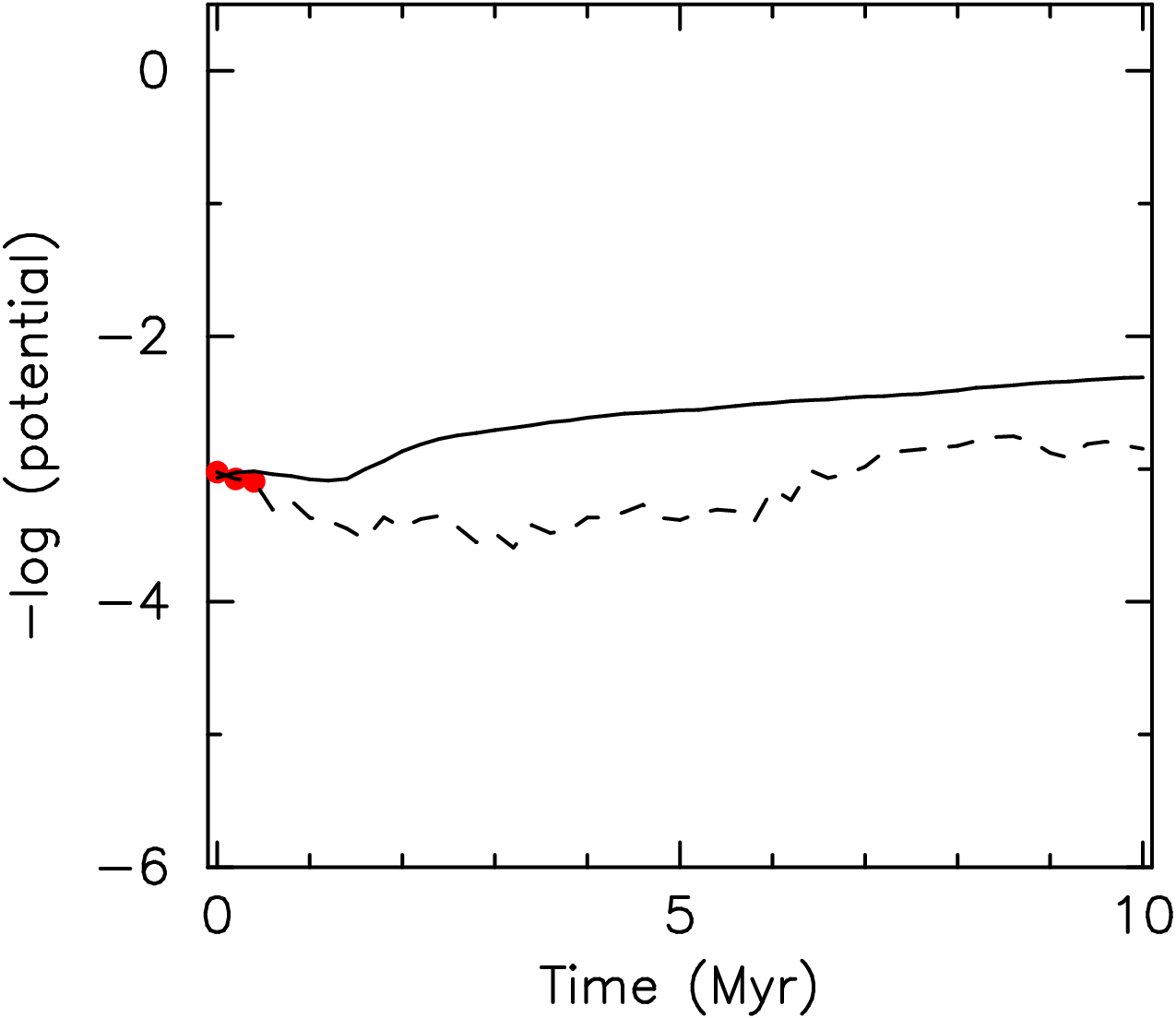}}}
\hspace*{0.3cm} 
\subfigure[Mass segregation evolution]{\label{subvirial-h}\rotatebox{270}{\includegraphics[scale=0.28]{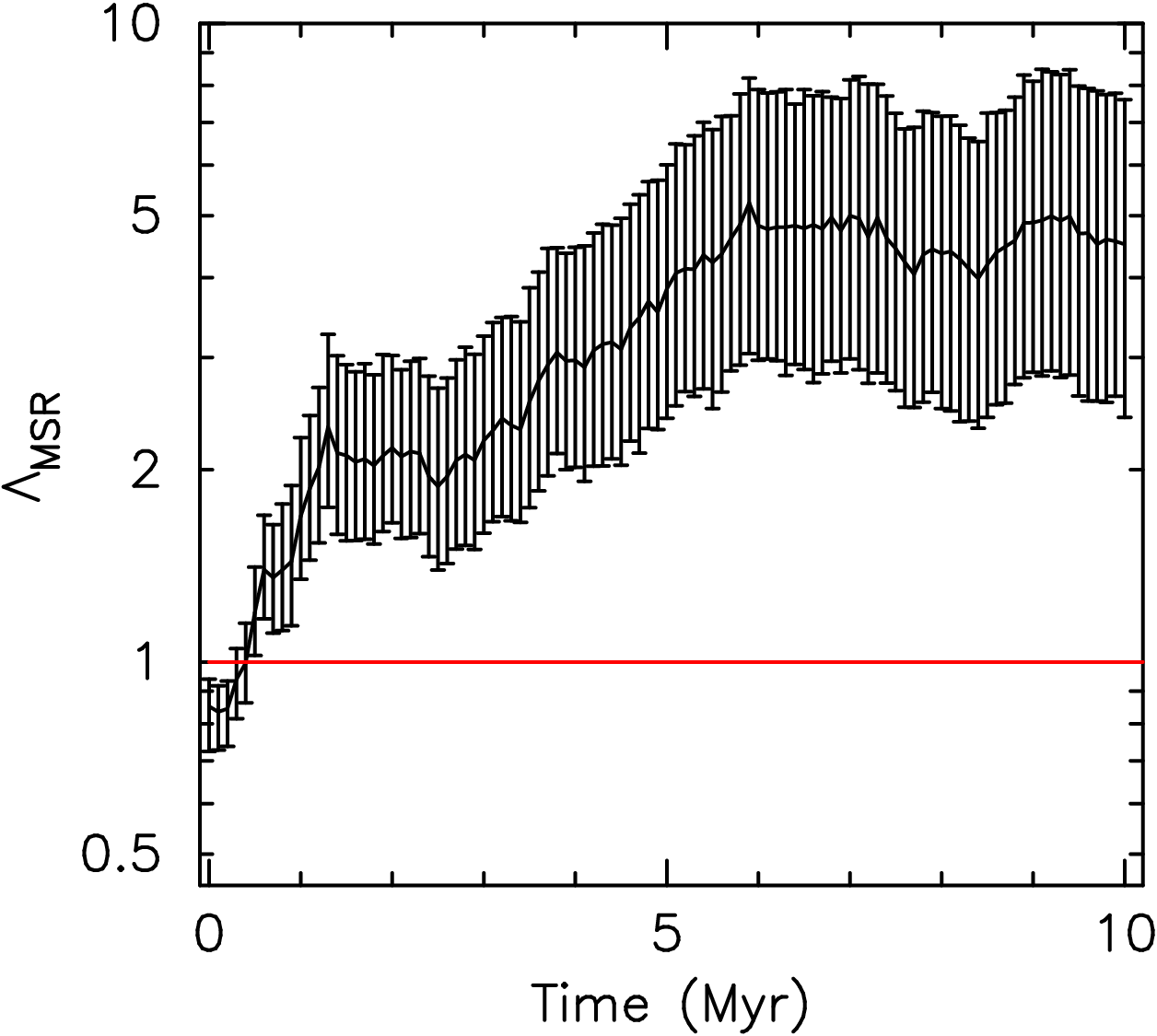}}} 
\hspace*{0.3cm}\subfigure[Local surface density evolution]{\label{subvirial-i}\rotatebox{270}{\includegraphics[scale=0.28]{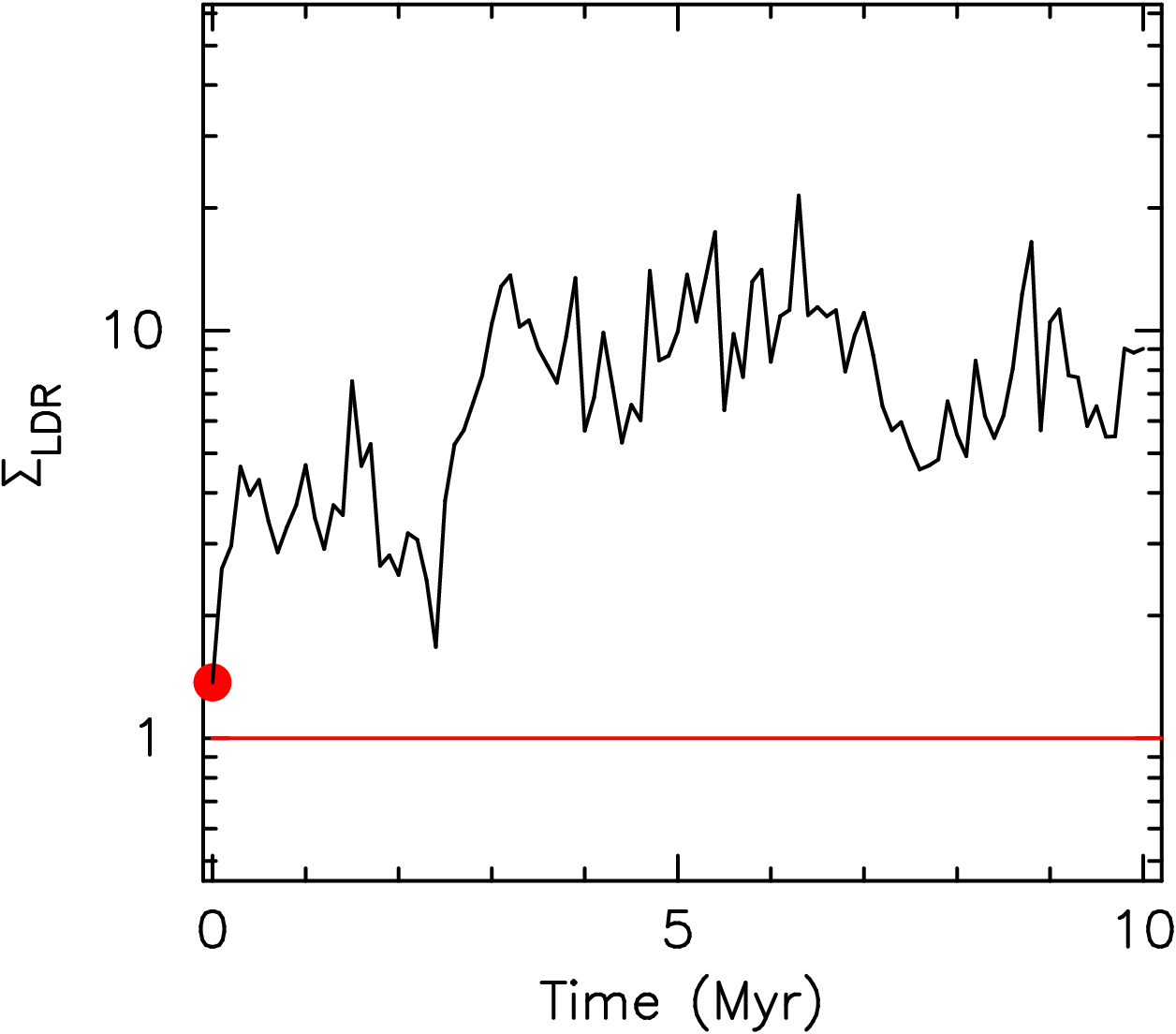}}}
\caption[bf]{Evolution of the potential for a subvirial, substructured $N$-body simulation. In panels (a - c) red triangles indicate the positions of the 10 most massive stars.  In panels (d-f) we show the potential as a function of stellar mass for each object in the star-forming region. The horizontal dashed lines show the median potential in the entire region, whereas the solid horizontal lines show the median potential of the ten most massive stars. Panel (g) shows the evolution of the median potential with time for all stars (the solid line) and the 10 most massive stars (dashed line). A red circle is plotted at the times where the difference between the potential of the massive stars and the lower-mass stars is \emph{not} significant. In panel (h) we show the evolution of mass segregation as quantified by the $\Lambda_{\rm MSR}$ parameter for the same simulation. In panel (i) we show the evolution of the local surface density ratio for massive stars, $\Sigma_{\rm LDR}$. Where the difference is not significant we plot a red circle.}
  \label{subvirial}
  \end{center}
\end{figure*}

\begin{figure*}
  \begin{center}
\setlength{\subfigcapskip}{10pt}

\hspace*{-1.5cm}\subfigure[0\,Myr]{\label{supervirial-a}\rotatebox{270}{\includegraphics[scale=0.28]{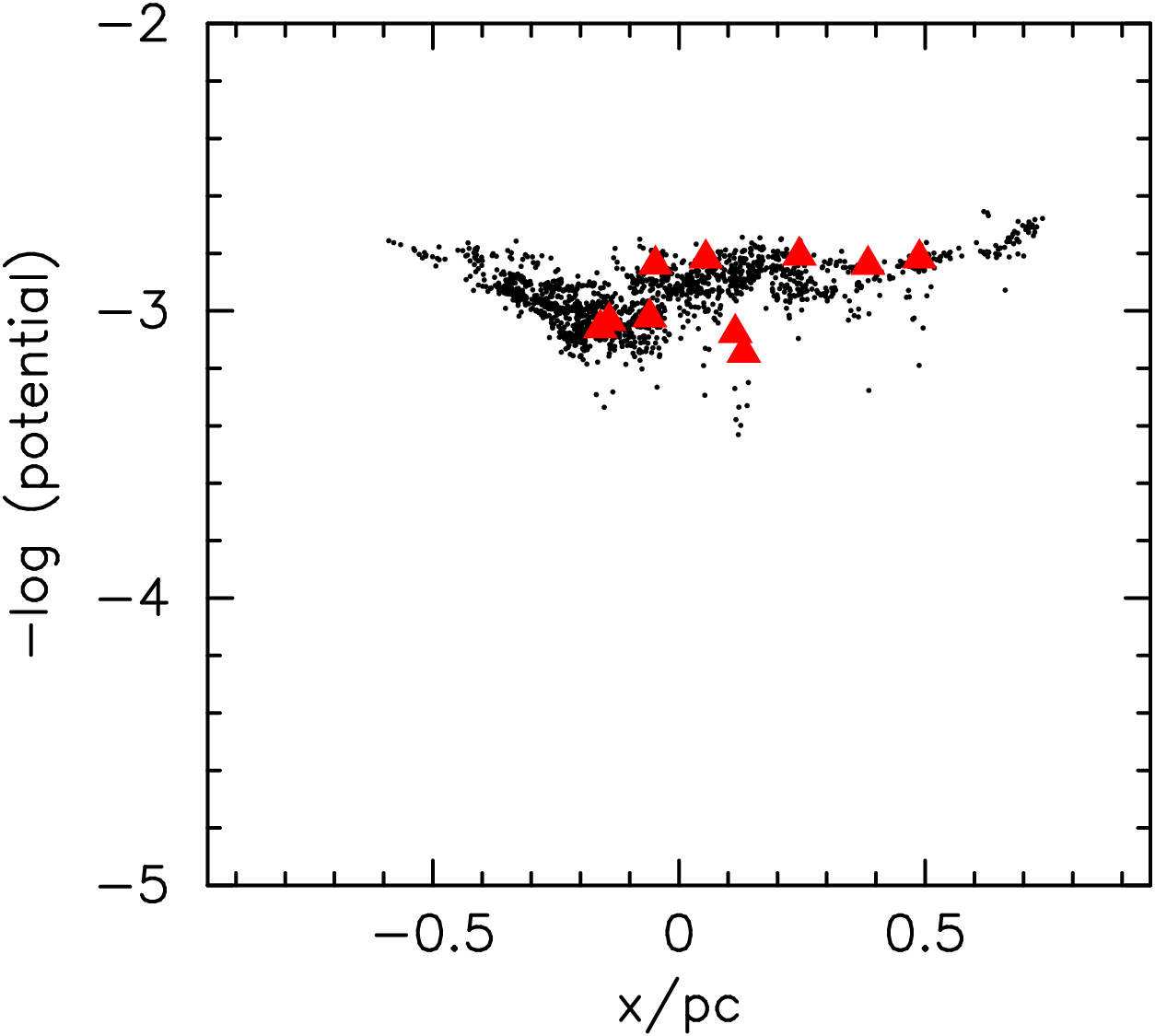}}}
\hspace*{0.3cm} 
\subfigure[4\,Myr]{\label{supervirial-b}\rotatebox{270}{\includegraphics[scale=0.28]{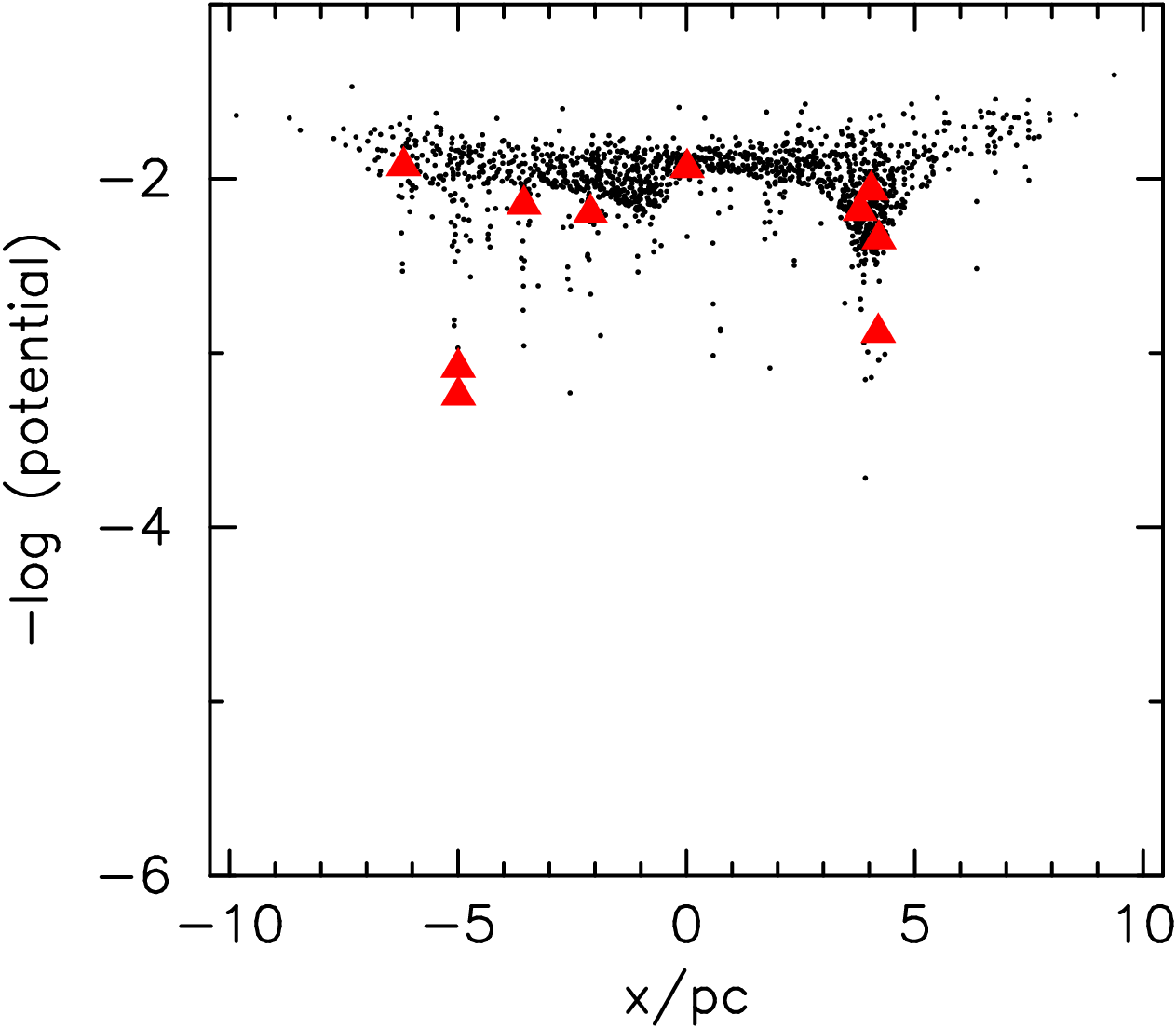}}} 
\hspace*{0.3cm}\subfigure[10\,Myr]{\label{supervirial-c}\rotatebox{270}{\includegraphics[scale=0.28]{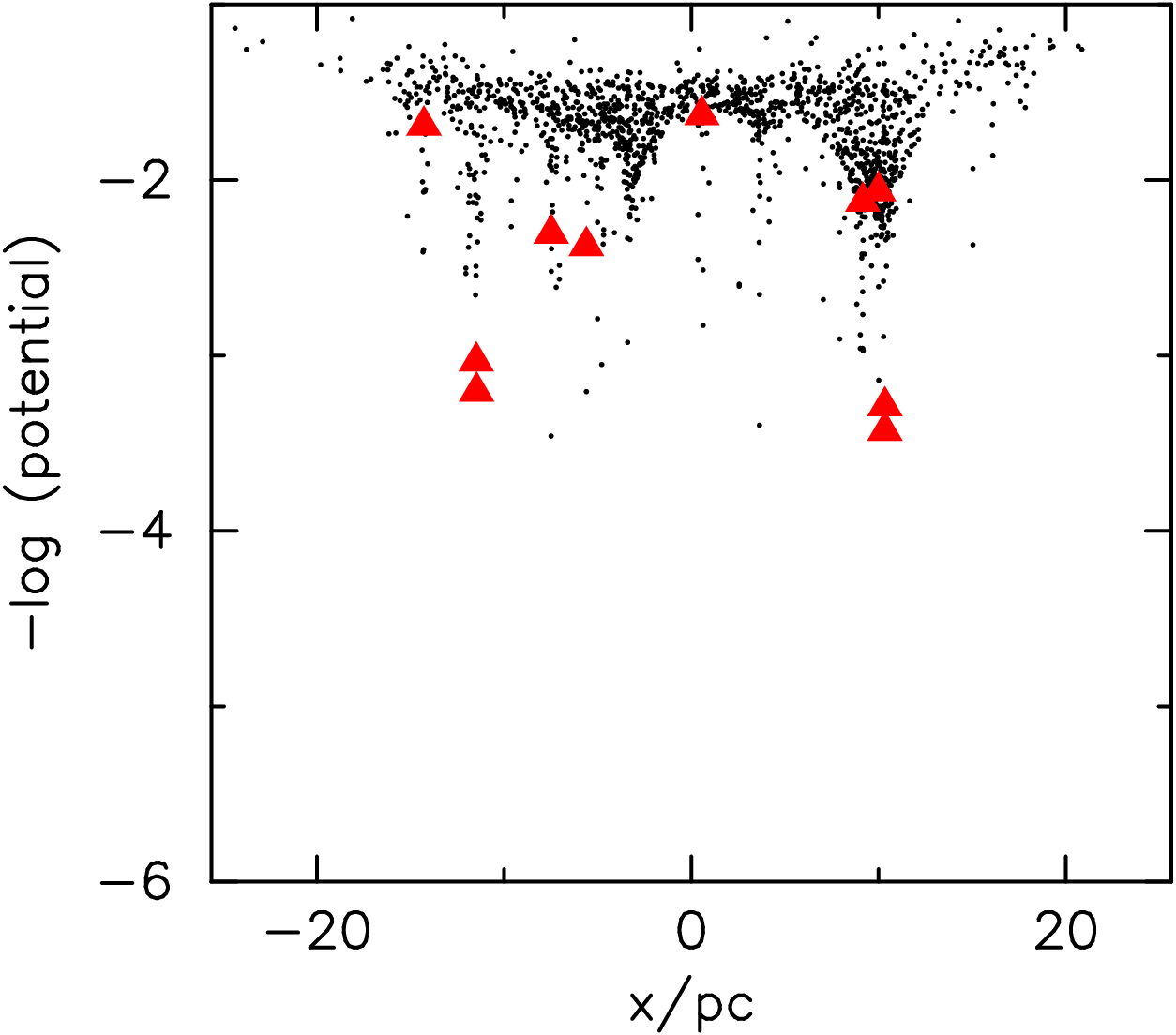}}}
\hspace*{-1.5cm}\subfigure[$\Phi - m$, 0\,Myr]{\label{supervirial-d}\rotatebox{270}{\includegraphics[scale=0.28]{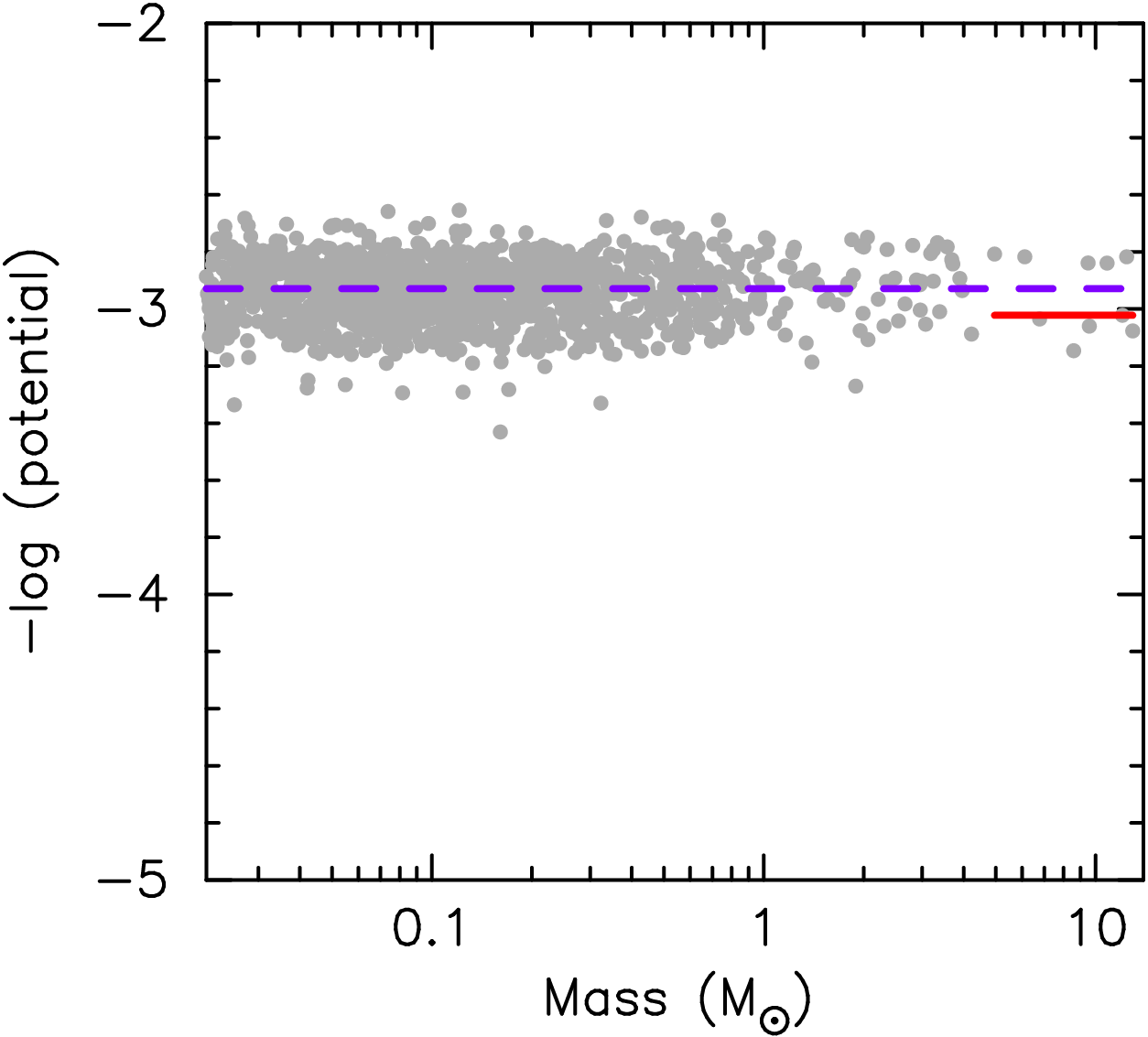}}}
\hspace*{0.3cm} 
\subfigure[$\Phi - m$, 4\,Myr]{\label{supervirial-e}\rotatebox{270}{\includegraphics[scale=0.28]{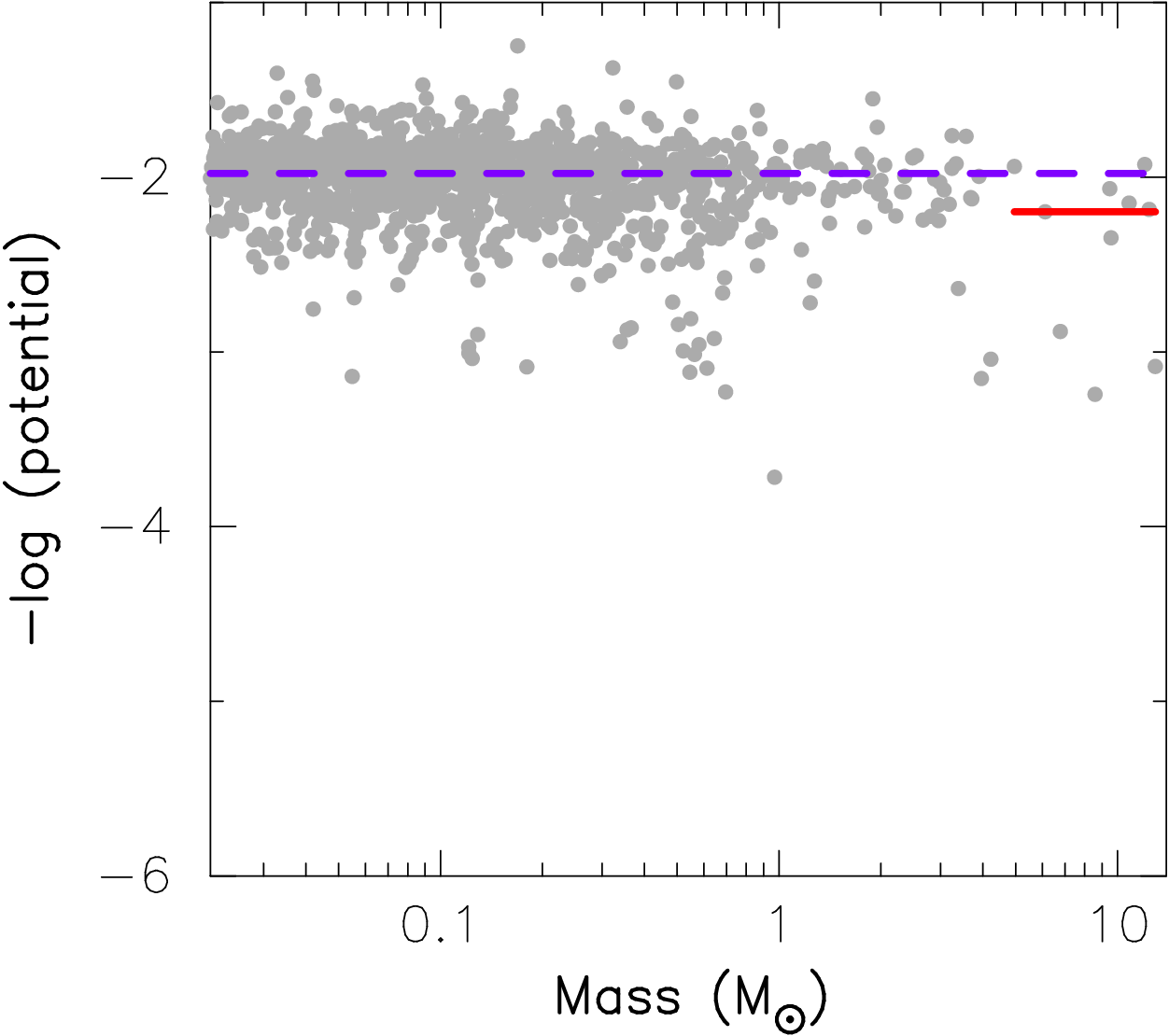}}} 
\hspace*{0.3cm}\subfigure[$\Phi - m$, 10\,Myr]{\label{supervirial-f}\rotatebox{270}{\includegraphics[scale=0.28]{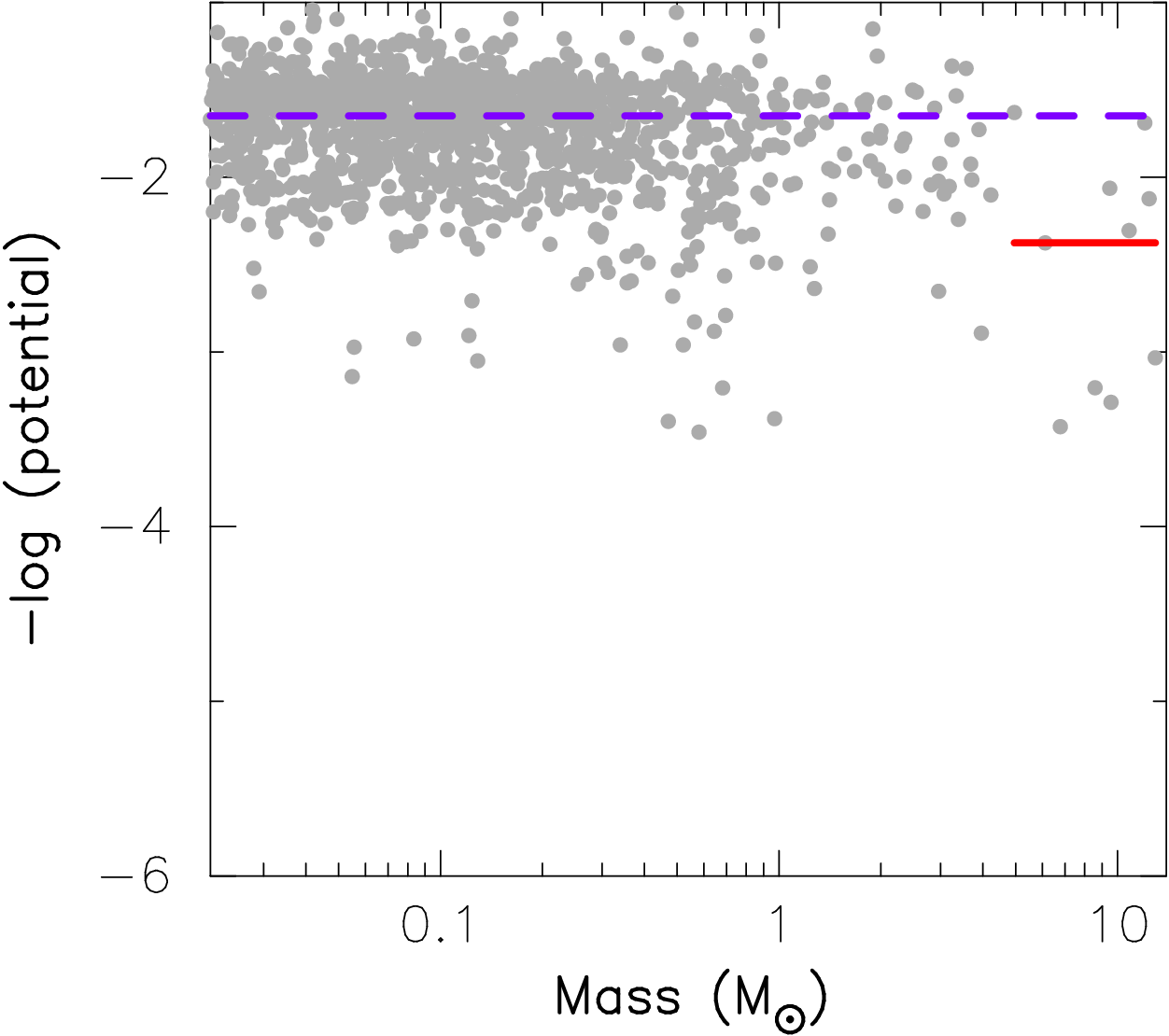}}}
\hspace*{-1.5cm}\subfigure[Evolution of the potential]{\label{supervirial-g}\rotatebox{270}{\includegraphics[scale=0.28]{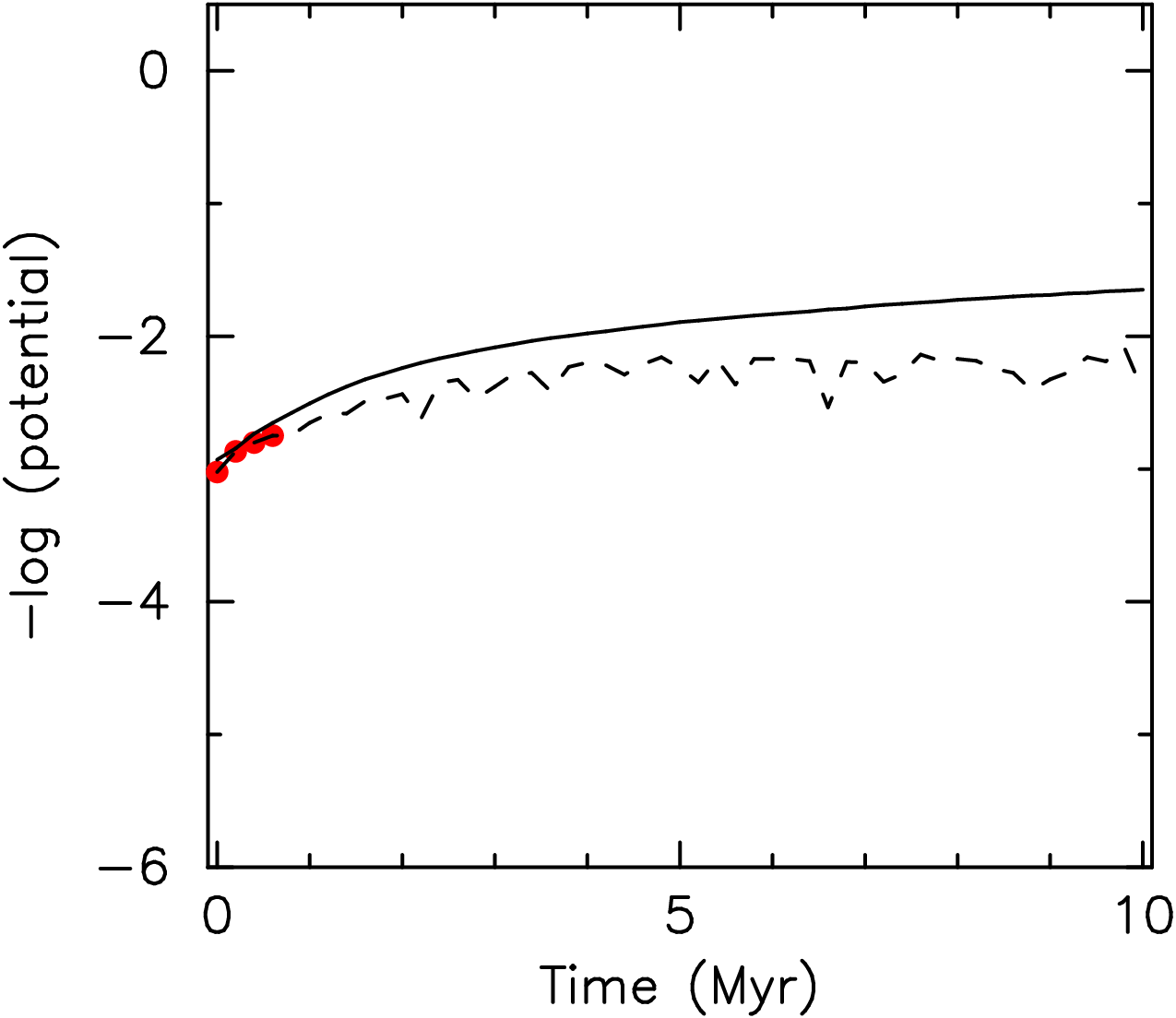}}}
\hspace*{0.3cm} 
\subfigure[Mass segregation evolution]{\label{supervirial-h}\rotatebox{270}{\includegraphics[scale=0.28]{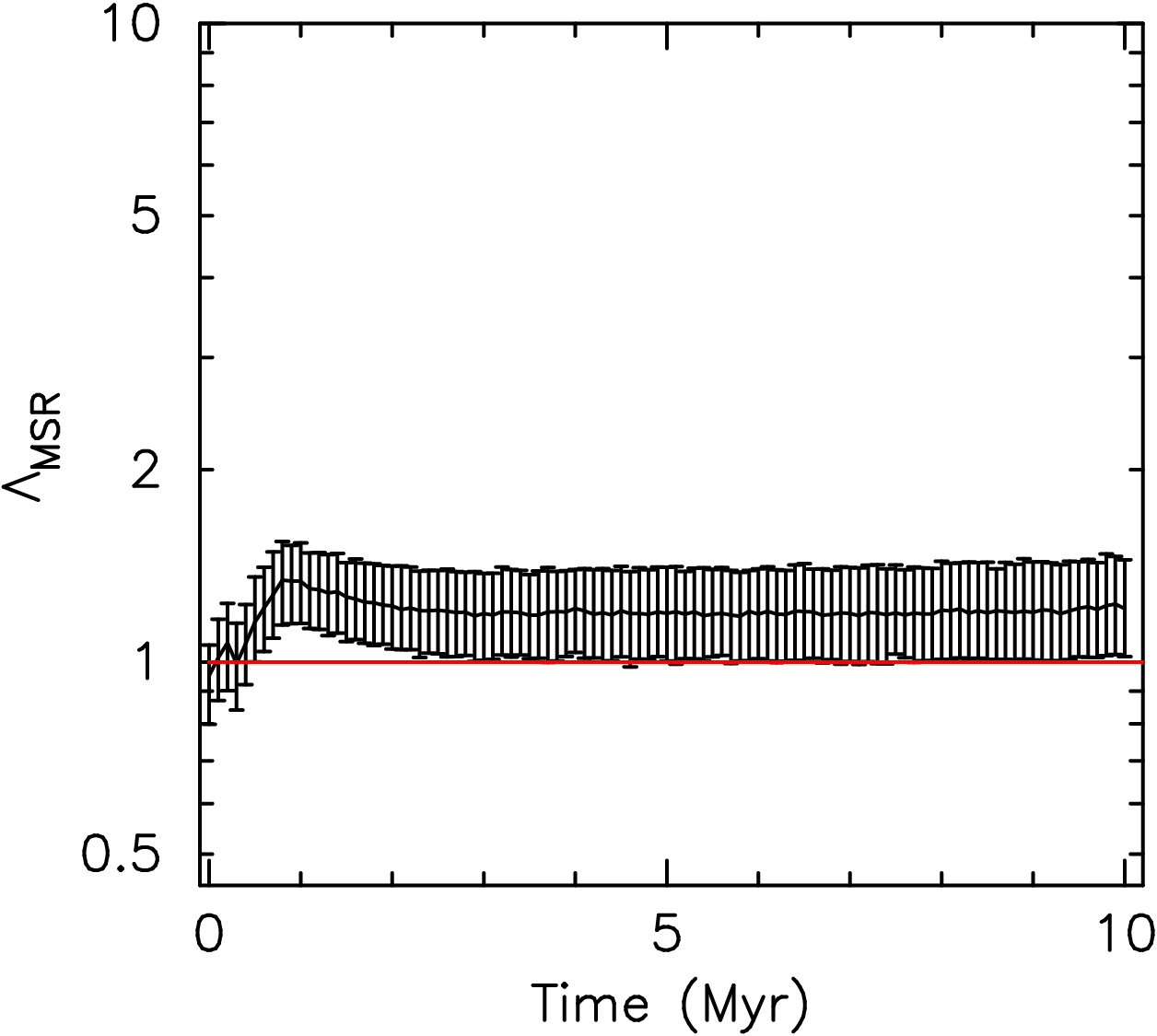}}} 
\hspace*{0.3cm}\subfigure[Local surface density evolution]{\label{supervirial-i}\rotatebox{270}{\includegraphics[scale=0.28]{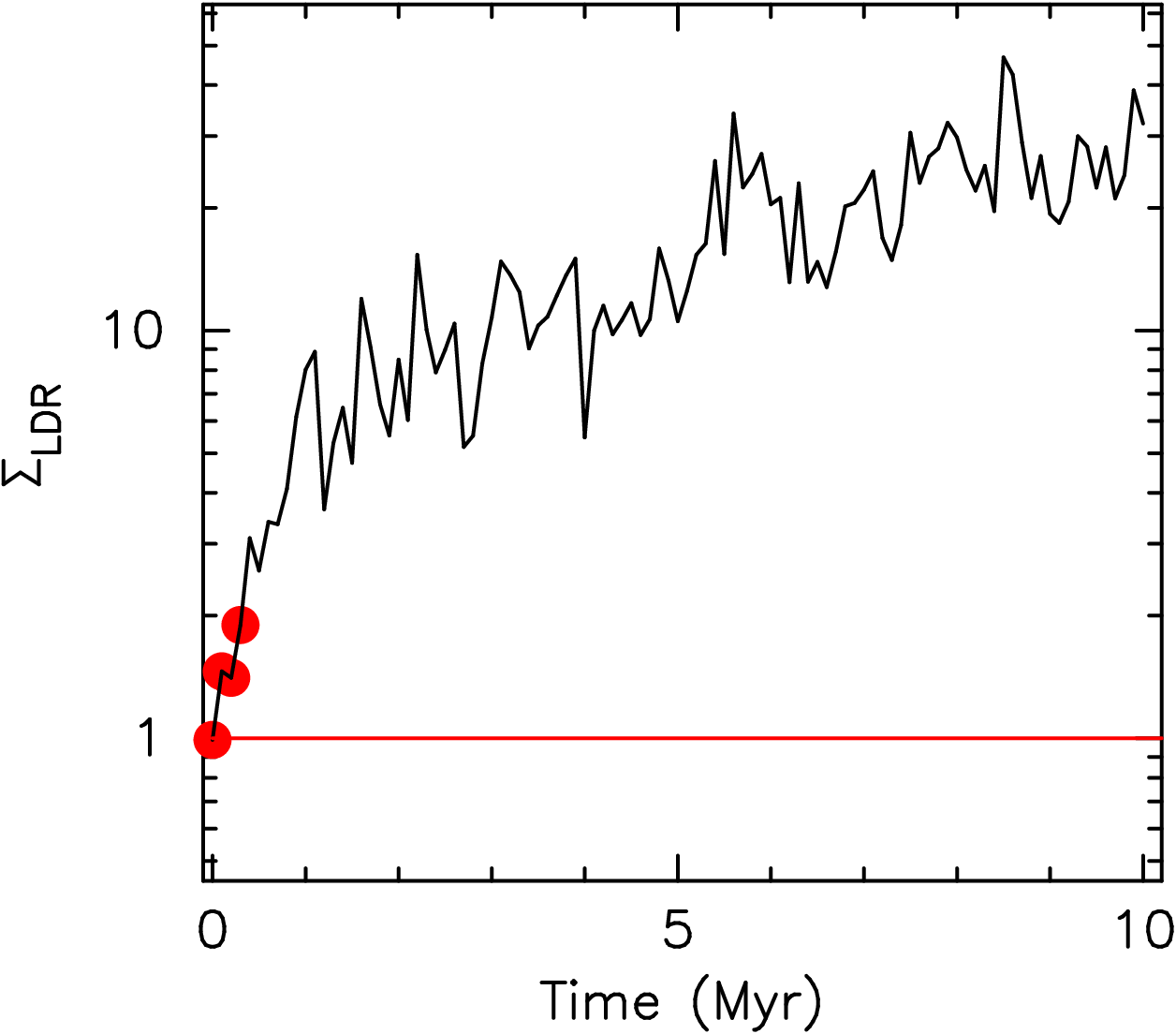}}}
\caption[bf]{Evolution of the potential for a supervirial, substructured simulation. In panels (a - c) red triangles indicate the positions of the 10 most massive stars. In panels (d-f) we show the potential as a function of stellar mass for each object in the star-forming region. The horizontal dashed lines show the median potential in the entire region, whereas the solid horizontal lines show the median potential of the ten most massive stars. Panel (g) shows the evolution of the median potential with time for all stars (the solid line) and the 10 most massive stars (dashed line). A red circle is plotted at the times where the difference between the potential of the massive stars and the lower-mass stars is \emph{not} significant. In panel (h) we show the evolution of mass segregation as quantified by the $\Lambda_{\rm MSR}$ parameter for the same simulation. In panel (i) we show the evolution of the local surface density ratio for massive stars, $\Sigma_{\rm LDR}$. Where the difference is not significant we plot a red circle.}
  \label{supervirial}
  \end{center}
\end{figure*}

\begin{table*}
\caption[bf]{Summary of the results from the SPH simulations and their subsequent dynamical evolution. The columns display the run ID, whether feedback was switched on, and the number of sink particles, $N_{\rm sinks}$,  the SPH mass resolution limit, $m_{\rm res}$, the total stellar (sink particle) mass,  $M_{\rm region}$, the median sink particle mass, $\tilde{m}_{\rm sink}$, then we show at 0, 2, and 10\,Myr whether the potential difference ratio $\Phi_{\rm PDR} >> 1$, the mass segregation ratio $\Lambda_{\rm MSR} >> 1$ and the local surface density ratio  $\Sigma_{\rm LDR} >> 1$. If these ratios do not significantly exceed unity, the entry is left blank.}
\begin{center}
\begin{tabular}{|c|c|c|c|c|c|c|c|c|c|}
\hline 
Run ID & Feedback &  $N_{\rm sinks}$ & $m_{\rm res}$ & $M_{\rm region}$ & $\tilde{m}_{\rm sink}$ & Time (Myr) & $\Phi_{\rm PDR} >> 1$? & $\Lambda_{\rm MSR} >> 1$?  & $\Sigma_{\rm LDR} >> 1$? \\
\hline
 &  & & & & & 0 & + & & + \\
 J & Off & 578 & 1\,M$_\odot$ & 3207\,M$_\odot$& 3.6\,M$_\odot$ & 2 & + & & \\
 &  &  & & & & 10 & + & & \\
\hline
 &  &  & & & & 0 & & + & \\
 J & Dual & 564 & 1\,M$_\odot$ & 2186\,M$_\odot$ & 2.1\,M$_\odot$ & 2 & + & & + \\
 &  &  & & & & 10 & + & & \\
\hline
\hline
 &  &  & & & & 0 & + & + & + \\
I  & Off & 186 & 1\,M$_\odot$ & 1271\,M$_\odot$ & 2.7\,M$_\odot$ & 2 & + & & \\
 &  &  & & & & 10 & + & + & + \\
\hline
 &  &  & & & & 0 & & & \\
 I & Dual & 132 & 1\,M$_\odot$ & 766\,M$_\odot$  & 3.0\,M$_\odot$ & 2  & & & \\
 &  &  & & & & 10 & + & & \\
\hline
\hline
 &  &  & & & & 0 & + & & \\
 UF & Off & 66 & 3\,M$_\odot$ & 1392\,M$_\odot$ & 4.8\,M$_\odot$ & 2 & + & &\\ 
 &  &  & & & & 10 & + & + & \\
\hline
 &  &  & & & & 0 & & & \\
 UF & Dual & 93 & 3\,M$_\odot$ & 841\,M$_\odot$ & 6.5\,M$_\odot$ & 2 & + & +  &\\
 &  &  & & & & 10 & + & & \\
\hline
\hline
 &  &  & & & & 0 & + & & + \\
 UP & Off & 340 & 3\,M$_\odot$ & 2718\,M$_\odot$ & 4.4\,M$_\odot$ & 2 & + & & \\
 &  &  & & & & 10 & & & \\
\hline
 &  &  & & & & 0 & + & &  \\
 UP & Dual & 343 & 3\,M$_\odot$ & 1926\,M$_\odot$ & 3.4\,M$_\odot$ & 2 & + & & \\
 &  &  & & & & 10 & + & & \\
\hline
\hline
 &  &  & & & &  0 & + & & + \\
 UQ & Off & 48 & 3\,M$_\odot$ & 723\,M$_\odot$ & 6.3\,M$_\odot$ & 2 & + & & \\ 
 &  &  & & & & 10 & & & \\
\hline
 &  &  & & & & 0 & + & & \\
 UQ & Dual & 77 & 3\,M$_\odot$ & 594\,M$_\odot$ & 4.7\,M$_\odot$ & 2 & & & \\
 &  &  & & & & 10 & + & + & \\
\hline
\hline
\end{tabular}
\end{center}
\label{sph_cluster_props}
\end{table*}

In Fig.~\ref{subvirial} we show the evolution of a subvirial star-forming region that collapses to form a cluster. We show the potential for each star as a function of distance from the centre of the region along the x-axis at 0\,Myr (Fig~\ref{subvirial-a}), 4\,Myr (Fig.~\ref{subvirial-b}) and 10\,Myr (Fig.~\ref{subvirial-c}). In all three panels, the red triangle symbols are the ten most massive stars in the simulation. 

Initially, the potential is rather uniform in that the most massive stars do not reside in a deeper potential well compared to low-mass stars. The initial substructure is then erased, and the potential becomes visibly smoother. As the cluster evolves, the potential well becomes shallower (i.e. the whole cluster becomes less strongly bound -- panels b and c), but the most massive stars visibly sit in a much deeper potential than the lower-mass stars.

We show the potential as a function of mass for every star in the simulation at 0, 4 and 10\,Myr in Figs.~\ref{subvirial-d},~\ref{subvirial-e}~and~\ref{subvirial-f}, respectively. In all panels, the median potential for the entire star-forming region is shown by the horizontal dashed line, and the median potential for the most massive stars is shown by the horizontal solid line. In this example, the massive stars initially are not in deeper potentials than average, but after 4\,Myr they are significantly deeper. The region as a whole evolves to a shallower potential due to the subsequent dynamical evolution, but the massive stars still reside in deeper potentials than average after 10\,Myr. 

In Fig.~\ref{subvirial-g} we show the evolution of the median potential for all stars (the solid line) and the median potential for the 10 most massive stars (the dashed line). Where the difference between these potentials is \emph{not} significant, we plot a red filled circle. A clear difference in potential for the most massive stars is apparent after the first 0.5\,Myr (the massive stars sit in a deeper potential).  The evolution of the mass segregation ratio, $\Lambda_{\rm MSR}$, is shown in Fig.~\ref{subvirial-h}. The region is not mass segregated initially, but violent relaxation leads to dynamical mass segregation within the first 5\,Myr. In Fig.~\ref{subvirial-i} we show the evolution of $\Sigma_{\rm LDR}$, which traces the local surface density around the most massive stars compared to the local surface density around all stars. The massive stars soon reside in areas of higher than average surface density (often because they are in the central regions of the cluster).

In summary, star-forming regions in cool collapse lead to the most massive stars sitting in deeper potentials than average, which would be observable as strong mass segregation, and high local surface density for the most massive stars. The results for the pure $N$-body models are summarised in Table~\ref{nbody_cluster_props}.

\subsection{Warm-expansion $N$-body simulation}

In Fig.~\ref{supervirial} we show the evolution of a supervirial star-forming region that expands to form an association. We show the potential for each star as a function of distance from the centre of the region along the x-axis at 0\,Myr (Fig~\ref{supervirial-a}), 4\,Myr (Fig.~\ref{supervirial-b}) and 10\,Myr (Fig.~\ref{supervirial-c}). In all three panels, the red triangle symbols are the ten most massive stars in the simulation. 

Because the potential is defined solely by the positions and masses of the stars with no dependence on velocity, the form of the initial potential is similar to the subvirial simulations presented in Fig.~\ref{subvirial}. Initially, the massive stars do not reside in deeper potentials than the lower-mass stars. As the region evolves, the initial substructure is partially erased and on average the stars reside in a shallower potential. Because the global motion of the region is expansion, the most massive stars evolve as quasi-isolated sub-systems, often dominating their immediate surroundings. This is apparent in Figs.~\ref{supervirial-b}~and~\ref{supervirial-c}, which display stalactite-like features in the potential, with the most massive stars residing at the base of each individual potential well.

We show the potential as a function of mass for every star in the simulation at 0, 4 and 10\,Myr in Figs.~\ref{subvirial-d},~\ref{subvirial-e}~and~\ref{subvirial-f}, respectively. In all panels, the median potential for the entire star-forming region is shown by the horizontal dashed line, and the median potential for the most massive stars is shown by the horizontal solid line. As in the subvirial regions, the massive stars initially are not in deeper potentials than average, but after 4\,Myr they are significantly deeper. The region as a whole also evolves to a shallower potential due to the subsequent dynamical expansion (albeit with less scatter than the subvirial star-forming regions), but the massive stars still reside in deeper potentials after 10\,Myr. 

In Fig.~\ref{supervirial-g} we show the evolution of the median potential for all stars (the solid line) and the median potential for the 10 most massive stars (the dashed line). Where the difference between these potentials is \emph{not} significant, we plot a red filled circle. After several Myr, the most massive stars in this supervirial simulation sit deeper in the potential than the average stars.  

In the evolution of the mass segregation ratio, $\Lambda_{\rm MSR}$, shown in Fig.~\ref{supervirial-h}, the most massive stars are not more concentrated than lower-mass stars, because the massive stars have not interacted with each other and formed centrally concentrated subsystems \citep[which readily occurs during cool-collapse,][]{Allison10,Parker14b,Parker16c}. In Fig.~\ref{supervirial-i} we show the evolution of $\Sigma_{\rm LDR}$, which traces the local surface density around the most massive stars compared to the local surface density around all stars. As the region evolves, the most massive stars sweep up retinues of low-mass stars, and have higher surface densities than low-mass stars.

As with the regions in cool collapse, in a supervirial, expanding region the most massive stars also migrate to sit in deeper potentials than the average stars. A spatial signature of this is that the most massive stars also reside in areas of high local surface density, although they are not mass segregated according to $\Lambda_{\rm MSR}$.

\subsection{SPH simulations}

\begin{figure*}
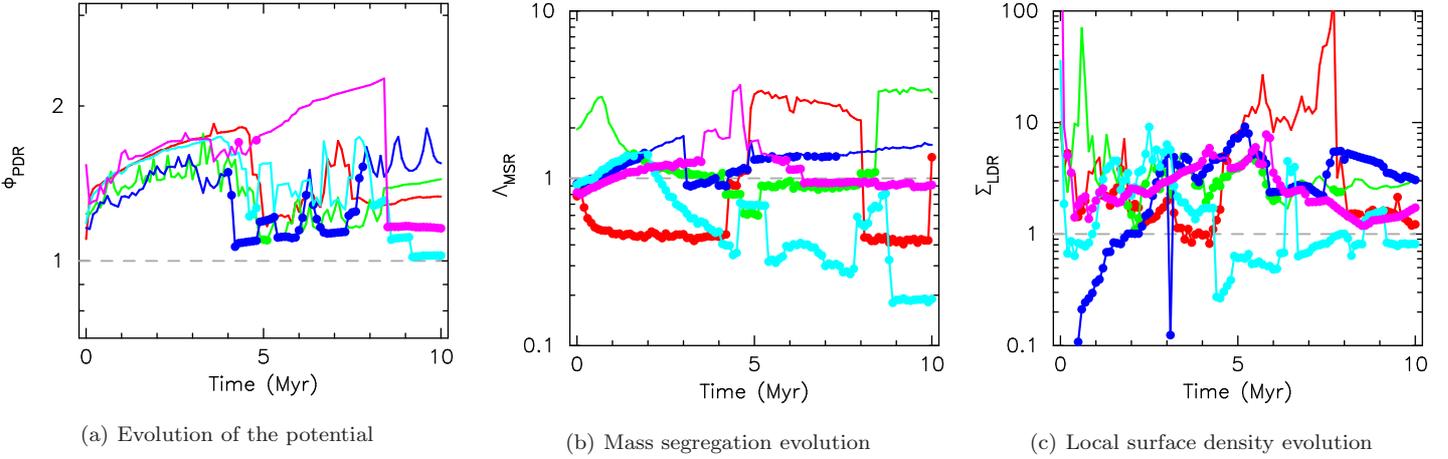

  \begin{center}
\setlength{\subfigcapskip}{10pt}

\hspace*{-1.5cm}\subfigure[Evolution of the potential]{\label{control-a}\rotatebox{270}{\includegraphics[scale=0.28]{Plot_JEDXTrig_N_Ovel__Potential_lines.ps}}}
\hspace*{0.3cm} 
\subfigure[Mass segregation evolution]{\label{control-b}\rotatebox{270}{\includegraphics[scale=0.28]{Plot_JEDXTrig_N_Ovel__Lambda_lines.ps}}} 
\hspace*{0.3cm}\subfigure[Local surface density evolution]{\label{control-c}\rotatebox{270}{\includegraphics[scale=0.28]{Plot_JEDXTrig_N_Ovel__Sigma_lines.ps}}}
\caption[bf]{Evolution of (a) the potential difference ratio, $\Phi_{\rm PDR}$, (b) the mass segregation ratio, $\Lambda_{\rm MSR}$, and (c) the local surface density ratio $\Sigma_{\rm LDR},$ in the $N$-body evolution from \citet{Parker15a}, which uses the simulations from \citet{Dale12a, Dale14} \emph{without feedback} switched on. A filled symbol is plotted where the deviation from unity is not significant (based on the numerical experiments in \citet{Parker15b} $\Lambda_{\rm MSR}<2\sigma$ from unity, and for $\Sigma_{\rm LDR}$ and $\Phi_{\rm PDR}$ KS test p-values $>$0.1 are insignificant departures from unity). The colours correspond to the following simulations: red -- run J; green -- run I; dark blue -- run UF; cyan -- run UP; magenta -- run UQ.}
\label{control}
  \end{center}
\end{figure*}

We now take the five SPH simulations published in \citet{Dale14}  and determine $\Phi_{\rm PDR}$, $\Lambda_{\rm MSR}$ and $\Sigma_{\rm LDR}$ for the sink particles at the end of the simulations. We then evolve the sink particles in the $N$-body integrator for a further 10\,Myr and determine $\Phi_{\rm PDR}$, $\Lambda_{\rm MSR}$ and $\Sigma_{\rm LDR}$ and present the values at 0\,Myr (after the SPH simulations are terminated), 2\,Myr and 10\,Myr in Table~\ref{sph_cluster_props}. Each simulation is represented by two rows in the table; one for the control run, and the other for the run with dual feedback mechanisms (photoionisation and stellar winds) switched on.

It is immediately apparent that the massive stars in the clusters generated by the SPH simulations are, simply going by the density of check marks in each column, likely to occupy deeper potentials than the low--mass stars (as traced by $\Phi_{\rm PDR}$), but unlikely either to be mass--segregated (as revealed by $\Lambda_{\rm MSR}$) or to occupy regions of higher--than average stellar surface density (as revealed by $\Sigma_{\rm LDR}$). With the exception of the cluster born in the Run I control simulation, the agreement of the three metrics is very poor, implying that they do not measure or trace the same properties of the stellar systems. $\Phi_{\rm PDR}$ stands out in particular, and it is evident from Table~\ref{sph_cluster_props} that massive stars can occupy substantially deeper potentials than low--mass stars \textit{without} being mass segregated or being in regions of unusually high stellar surface density.


\subsubsection{The control runs}

\begin{figure*}
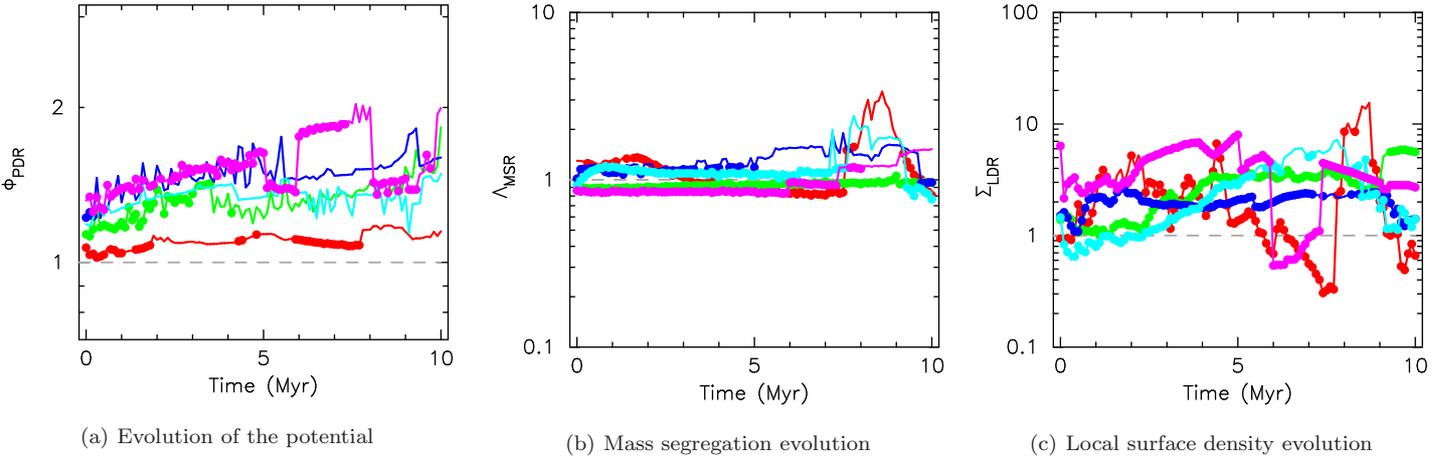

  \begin{center}
\setlength{\subfigcapskip}{10pt}

\hspace*{-1.5cm}\subfigure[Evolution of the potential]{\label{dual-a}\rotatebox{270}{\includegraphics[scale=0.28]{Plot_JEDXTrig_B_Ovel__Potential_lines.ps}}}
\hspace*{0.3cm} 
\subfigure[Mass segregation evolution]{\label{dual-b}\rotatebox{270}{\includegraphics[scale=0.28]{Plot_JEDXTrig_B_Ovel__Lambda_lines.ps}}} 
\hspace*{0.3cm}\subfigure[Local surface density evolution]{\label{dual-c}\rotatebox{270}{\includegraphics[scale=0.28]{Plot_JEDXTrig_B_Ovel__Sigma_lines.ps}}}
\caption[bf]{Evolution of (a) the potential difference ratio, $\Phi_{\rm PDR}$, (b) the mass segregation ratio, $\Lambda_{\rm MSR}$, and (c) the local surface density ratio $\Sigma_{\rm LDR},$ in the $N$-body evolution from \citet{Parker15a}, which uses the simulations from \citet{Dale12a, Dale14} with \emph{dual feedback} switched on. A filled symbol is plotted where the deviation from unity is not significant (based on the numerical experiments in \citet{Parker15b} $\Lambda_{\rm MSR}<2\sigma$ from unity, and for $\Sigma_{\rm LDR}$ and $\Phi_{\rm PDR}$ KS test p-values $>$0.1 are insignificant departures from unity). The colours correspond to the following simulations: red -- run J; green -- run I; dark blue -- run UF; cyan -- run UP; magenta -- run UQ. }
\label{dual}
  \end{center}
\end{figure*}

The evolution of the control run simulations is shown in Fig.~\ref{control}. In all control run simulations, the massive stars reside in deeper stellar potentials than the average stars at the end of the SPH simulation, but before the subsequent $N$-body evolution. In one simulation (Run I -- the green line) the massive stars are significantly deeper within the potential, display marginal mass segregation according to $\Lambda_{\rm MSR}$ and the massive stars are in areas of higher surface density according to $\Sigma_{\rm LDR}$. However, subsequent dynamical and stellar evolution cause the significant $\Lambda_{\rm MSR}$ and  $\Sigma_{\rm LDR}$ signals to be erased after $\sim$2\,Myr, before stellar and dynamical evolution leads to a re-emergence of mass segregation and high relative stellar surface densities after 8.5\,Myr. It is unclear if the spatial distribution from star formation in this simulation would be observable, given that it is erased so early on.

In the remaining simulations, the massive stars reside in areas of higher than average surface density at the end of the SPH calculations in Runs J, UP and UQ (the red, cyan and magenta lines, respectively). However, none of them display mass segregation according to $\Lambda_{\rm MSR}$. Subsequent dynamical and stellar evolution also leads to the presence (and absence) of a different spatial distribution for the most massive stars (either in  $\Lambda_{\rm MSR}$ or $\Sigma_{\rm LDR}$), suggesting that the spatial signature of the initial conditions is rarely preserved until these regions can be observed (at ages $>$1\,Myr).


\subsubsection{Dual feedback runs}

The evolution of the dual-feedback simulations is shown in Fig.~\ref{dual}. In simulations Run UP (the cyan line) and UQ (magenta line) the massive stars sit in a deeper potential than the average stars. However, the subsequent dynamical evolution of Run UQ erases this signature. The surface density ratio $\Sigma_{\rm LDR}$ does not exceed unity at the start of any of the simulations and subsequent dynamical evolution tends not to lead to high $\Sigma_{\rm LDR}$ values. The only simulation that shows strong mass segregation ($\Lambda_{\rm MSR} > 2$) is Run J (the red line) between 8 and 9\,Myr. 

The simulations demonstrate that the presence of feedback significantly reduces the differences in the spatial distributions of the most massive stars with respect to the lower-mass stars, compared to the control runs where feedback is not enabled.\\

The evolution of $\Phi_{\rm PDR}$, $\Lambda_{\rm MSR}$ and $\Sigma_{\rm LDR}$ is much more noisy in the control run simulations. In particular, in runs~J (red lines) and UF (dark blue lines) there are significant changes in all three diagnostics between 3 and 5\,Myr. This is because the control runs have a somewhat top-heavy IMF due to the runaway growth of massive stars that is not regulated by feedback \citep{Dale12a,Dale14}. These simulations each contain between five and ten stars with initial masses $>$40\,M$_\odot$ (control run~UF has four stars with initial masses $>$90\,M$_\odot$), which rapidly evolve within 5\,Myr. As these stars lose mass through their individual stellar evolution, they drop out of the subset of the ten most massive stars and so the diagnostics change rapidly. This merely serves to highlight one of the many challenges of quantifying mass segregation; stars not only move, but the $X$ most massive stars in a cluster of a given age may not correspond to the $X$ most massive stars at formation. 

\begin{figure*}
  \begin{center}
    {\includegraphics[scale=0.5]{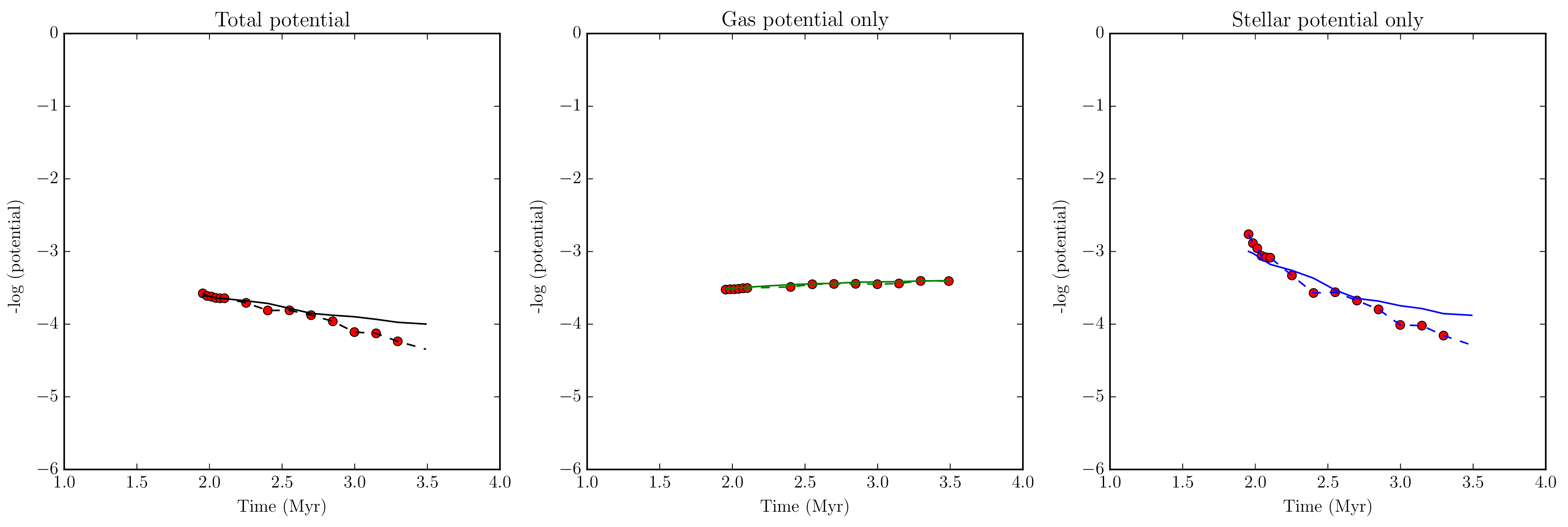}}
\end{center}
\caption[bf]{Evolution of the median potentials of the ten most massive stars (dashed lines) and of all stars (solid lines) in the SPH control run J, in the case where the total potential (left panel), the gas potential (centre panel) or the stellar potential (right panel) are computed. Where the difference between these values is not significant, we plot a filled red circle.}
\label{sph}
\end{figure*}

\begin{figure*}
  \begin{center}
    {\includegraphics[scale=0.5]{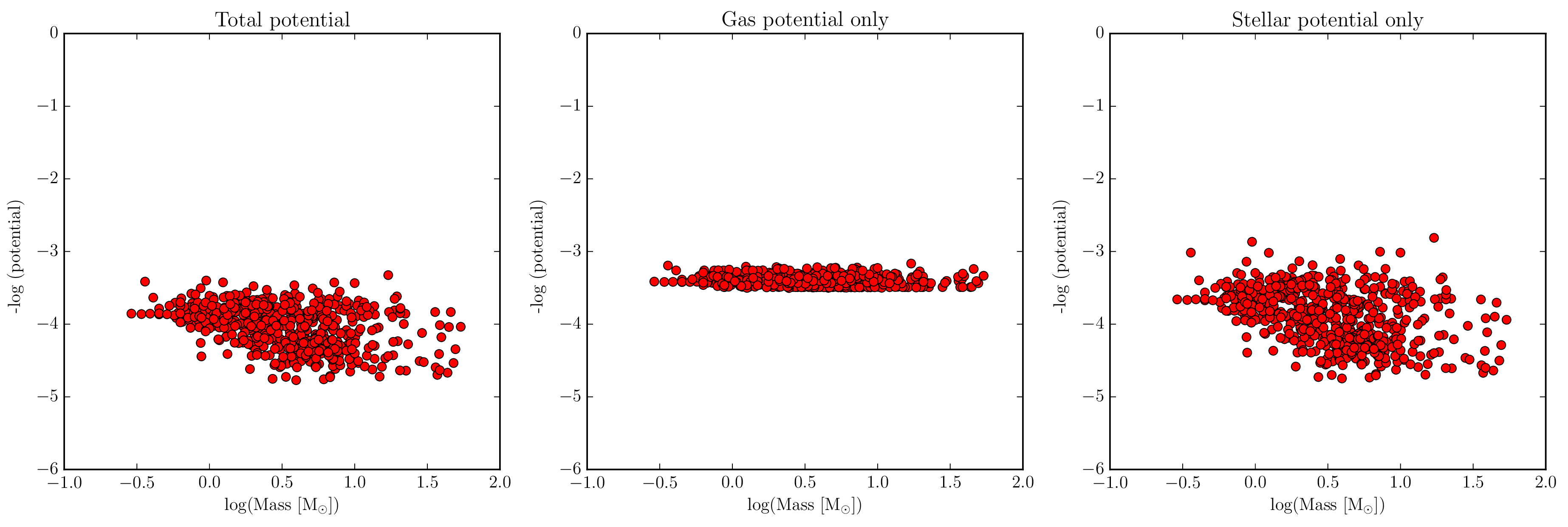}}
\end{center}
\caption[bf]{Potential versus mass at  the end-point (3.49\,Myr) of the SPH control run J for the total potential (left panel), the gas potential (centre panel) or the stellar potential (right panel).}
\label{sph:phi-mass}
\end{figure*}

\subsubsection{Evolution of the stellar potentials during star formation}

In the SPH control run simulations where no feedback is enabled, the most massive stars end up in significantly deeper potentials than the lower mass stars in the clusters. In Fig.~\ref{sph} we show the evolution of the potential in a typical simulation (Run~J). The massive stars end up in significantly deeper potentials than the low mass stars only towards the end of the SPH simulations, suggesting that this is a stellar dynamical effect rather than something related to the star formation process itself. This is further corroborated by the centre panel of Fig.~\ref{sph}, which shows that, for much of the simulation, the massive stars do not sit in deeper regions of \textit{gas} potential, but that they do occupy regions of deeper \textit{stellar potential}. If we examine the potential as a function of sink-particle mass at the end-point of this SPH simulation (3.49\,Myr, Fig.~\ref{sph:phi-mass}), we clearly see that the total potential (panel a) is dominated by the potential of the stars, rather than the gas.  This is a consequence of the findings of \citet{Dale15b}, who show that the stars in the SPH simulations very soon find themselves in stellar--dominated potentials.

\section{Discussion}

The first goal of our analysis was to understand how the potential of massive stars with respect to lower-mass stars in a star-forming region relates to the spatial distribution. The pure $N$-body simulations discussed in Section 3.1 have very dense ($>10^3$\,M$_\odot$\,pc$^{-3}$) initial conditions, and undergo extensive dynamical evolution in the first 10\,Myr. 

In simulations of subvirial regions undergoing cool collapse, the most massive stars rapidly sink into deeper potentials than the lower-mass stars, and because the massive stars end up in the centre of the cluster, strong signatures of both mass segregation ($\Lambda_{\rm MSR} >> 1$) and higher surface densities than the lower-mass stars ($\Sigma_{\rm LDR} >>1$) occur. For these collapsing star-forming regions, the three metrics we use to quantify the relative spatial distribution of the most massive stars are all in agreement. 

However, \citet{Parker15b} show that $\Lambda_{\rm MSR}$ and $\Sigma_{\rm LDR}$ do not always trace the same spatial distributions in star-forming regions with non-spherical geometries. This is apparent in our $N$-body simulations of star-forming regions that are initially supervirial, i.e.\,\,expanding. The massive stars dominate their local environment, which leads to them sitting in deeper potentials compared to low-mass stars. By sitting in deeper potentials, the massive stars can acquire retinues of low-mass stars (they are effectively accreting sub-clusters) and this is shown by the $\Sigma_{\rm LDR}$ ratio, which increases as the simulation progresses. In these supervirial simulations the massive stars rarely interact with \emph{each other}, and the cluster therefore does not become mass segregated. These results immediately demonstrate that the most massive stars can sit in deeper potential wells, but not display mass segregation \citep[either in $\Lambda_{\rm MSR}$, or in a similar tracer of central concentration, such as comparing radial mass functions of subsets of stars -- see e.g.][]{Hillenbrand98,Gouliermis04,Littlefair04,Parker15b}.  

In our SPH control run simulations of star formation, the massive stars usually sit in deeper potential wells than the lower-mass stars, and this is often reflected by the high relative surface density at the ends of the SPH simulations, and therefore the beginning of the $N$-body calculations. However, stellar and dynamical evolution generally erases this signature in $<$2\,Myr. These regions are spatially substructured \citep{Parker13a}, and do not display prominent mass segregation according to $\Lambda_{\rm MSR}$.


In the control simulations, accretion by the massive stars is entirely unregulated by feedback and they are able to grow to large masses, in excess of 100\,M$_{\odot}$ in some cases. These very large masses lead to large values of both $\Phi_{\rm PDR}$ and $\Sigma_{\rm LDR}$, as the very massive stars occupy deep potentials into which gas is continuously funnelled and where large numbers of low--mass stars are able to form.

In the SPH simulations that also include feedback from winds and photoionisation, the massive stars generally do not sit in deeper potentials. In these calculations, feedback disrupts the flows delivering gas into the clouds' main potential wells, clearing the clusters of gas. This has the effect of arresting the growth of the most massive stars, making the cluster potentials shallower with respect to those in the control simulations, and redistributing stars and star formation over larger volumes. This reduces $\Phi_{\rm PDR}$ as well as $\Sigma_{\rm LDR}$ \citep{Parker13a,Parker15a}. In simulations that include magnetic fields \citep[e.g.][]{Myers14}, high $\Sigma_{\rm LDR}$ values are reported, which may be due to the magnetic fields constraining gas in rather small star-forming volumes. \citet{Myers14} do not report on whether the massive stars sit deeper in the potential, but as we have seen, a high  $\Phi_{\rm PDR}$ ratio will often (but not always) present itself in tandem with a high $\Sigma_{\rm LDR}$ value. 

It is clear from our results that $\Phi_{\rm PDR}$,  $\Lambda_{\rm MSR}$ and $\Sigma_{\rm LDR}$ are measuring very different properties of the spatial distributions of the most massive stars. Typically,  $\Phi_{\rm PDR}$ is significantly higher than unity more often than $\Sigma_{\rm LDR}$ , which in turn is significantly higher than unity more often than $\Lambda_{\rm MSR}$. We interpret this as being due to different levels of interaction of the most massive stars with both each other, and their surroundings. The occurrence of massive stars in deeper potentials than average is either driven by significant accretion of gas in the SPH simulations (which is not always directly related to interactions with other stars), or by the massive stars dominating their local environment in the $N$-body simulations. A significant  $\Sigma_{\rm LDR}$ ratio only occurs if the massive stars are dominating their environment in the regime of stellar interactions. Finally, a high $\Lambda_{\rm MSR}$ usually only occurs if the massive stars have interacted with \emph{each other}.  

 
In the simulations we present here, massive stars do form from competitive accretion, but this does not always translate into an observable difference in their spatial distribution compared to lower-mass stars, even in the purely hydrodynamical control runs. When extra physics such as feedback is included, the massive stars are less likely to reside in deeper potentials, and consequently display even less of a spatial signature of competitive accretion than in the control runs. 

Following star formation, dynamical interactions alter the initial spatial distribution, often on rapid timescales \citep[see also][]{McMillan07,Allison09b,Moeckel09b,Allison10,Olczak11,Parker14b}, which further confuses the interpretation of mass segregation or enhanced surface density in young star-forming regions and clusters. Recently, \citet{Kuznetsova15} have shown that mass segregation is accelerated even during the star formation process if the gas from which stars form is subvirial.

Our intepretation of these results is that a spatial signature of competitive accretion would not be observable in a young star-forming region, even if it plays a dominant role in the formation of massive stars. We emphasize that our results do not preclude competitive accretion (or indeed monolithic collapse) from being the main formation channel for massive stars; rather, any spatial signature it imprints in the distribution of massive stars either does not exist, or cannot be identified with current methods to any statistical significance. 





\section{Conclusions}

We analyse $N$-body simulations of the evolution of star-forming regions and SPH simulations of star formation to look for differences in the spatial distribution of the most massive stars compared to lower-mass stars. In particular, we examine whether the most massive stars that form from competitive accretion sit in deeper potentials, and whether this translates into a different observable spatial distribution. Our conclusions are the following:

(i) In $N$-body simulations of subvirial (collapsing) star-forming regions, the massive stars move to deeper potentials, are more centrally concentrated, and have higher surface densities than the lower-mass stars in the region. In supervirial (expanding) simulations, the massive stars move to deeper potentials, and also collect retinues of lower-mass stars so their relative surface densities are higher than the average in the star-forming region. However, because the massive stars rarely interact with each other in these expanding regions, they do not mass segregate.

(ii) In purely hydrodynamical SPH simulations of star-formation, the most massive stars tend to reside in deeper potential wells than the lower-mass stars following star formation. This often (but not always) translates into high relative surface densities for the most massive stars. 

(iii) When feedback from photoionisation and stellar winds is included in the SPH simulations, massive stars can form via competitive accretion, but their formation is regulated by the feedback and the runaway growth of massive stars does not occur. The net result of this is that fewer simulations display high differences in the potential and local surface density of massive stars compared to lower-mass stars.

(iv) Massive stars can subsequently move into deeper potentials, attain high relative surface densities and mass segregate as the star-forming regions dynamically evolve. 

(v) Our results suggest that primordial mass segregation, or any other preferential spatial distribution of massive stars, should not be expected from the competitive accretion formation mechanism. We emphasize that this does not preclude competitive accretion as a viable formation channel for massive stars, but rather that no observable signature would be expected.








\section*{Acknowledgments}

We thank the anonymous referee for their comments and suggestions. RJP acknowledges support from the Royal Society in the form of a Dorothy Hodgkin Fellowship. 

\bibliography{general_ref}

\label{lastpage}

\end{document}